\documentclass[12pt]{article}
\usepackage{latexsym}
\usepackage{epsfig,amssymb,euscript,slashed}
\usepackage[linktocpage]{hyperref}
\usepackage{amsmath}
\usepackage{cite} 
\usepackage{array,calc,epsfig}
\usepackage{bbm}
\usepackage{fancybox}

\def\ii {{\rm i}}

\numberwithin{equation}{section} 
\usepackage[]{graphicx}
\usepackage{color}

\begin{document}
\font\cmss=cmss10 \font\cmsss=cmss10 at 7pt

\begin{flushright}{
\scriptsize QMUL-PH-17-21}
\end{flushright}
\hfill
\vspace{18pt}
\begin{center}
{\Large 
\textbf{Unitary 4-point correlators from classical geometries
}}

\end{center}

\vspace{8pt}
\begin{center}
{\textsl{Alessandro Bombini$^{\,a,b}$, Andrea Galliani$^{\,a, b}$, Stefano Giusto$^{\,a, b}$, Emanuele Moscato$^{\,c}$ and Rodolfo Russo$^{\,c}$}}

\vspace{1cm}

\textit{\small ${}^a$ Dipartimento di Fisica ed Astronomia ``Galileo Galilei",  Universit\`a di Padova,\\Via Marzolo 8, 35131 Padova, Italy} \\  \vspace{6pt}

\textit{\small ${}^b$ I.N.F.N. Sezione di Padova,
Via Marzolo 8, 35131 Padova, Italy}\\
\vspace{6pt}

\textit{\small ${}^c$ Centre for Research in String Theory, School of Physics and Astronomy\\
Queen Mary University of London,
Mile End Road, London, E1 4NS,
United Kingdom}\\
\vspace{6pt}

\end{center}

\vspace{12pt}

\begin{center}
\textbf{Abstract}
\end{center}

\vspace{4pt} {\small
\noindent 
We compute correlators of two heavy and two light operators in the strong coupling and large $c$ limit of the D1D5 CFT which is dual to weakly coupled AdS$_3$ gravity. The light operators have dimension two and are scalar descendants of the chiral primaries considered in arXiv:1705.09250, while the heavy operators belong to an ensemble of Ramond-Ramond ground states. We derive a general expression for these correlators when the heavy states in the ensemble are close to the maximally spinning ground state. For a particular family of heavy states we also provide a result valid for any value of the spin. In all cases we find that the correlators depend non-trivially on the CFT moduli and are not determined by the symmetries of the theory, however they have the properties expected for correlators among pure states in a unitary theory, in particular they do not decay at large Lorentzian times.}

\vspace{1cm}

\thispagestyle{empty}

\vfill
\vskip 5.mm
\hrule width 5.cm
\vskip 2.mm
{
\noindent  {\scriptsize e-mails:  {\tt alessandro.bombini@pd.infn.it, andrea.galliani@pd.infn.it, stefano.giusto@pd.infn.it, e.moscato@qmul.ac.uk, r.russo@qmul.ac.uk} }
}

\setcounter{footnote}{0}
\setcounter{page}{0}

\newpage


\section{Introduction}
\label{sec:intro}

In the AdS/CFT context black holes are dual to ensembles of ``heavy'' CFT states whose conformal dimension scales as the central charge. A prototypical case is that of the Strominger-Vafa~\cite{Strominger:1996sh} black hole which admits an AdS$_3 \times S^3$ decoupling limit and a dual description in terms of a 2-dimensional SCFT~\cite{Maldacena:1997re} often dubbed D1D5 CFT. The key breakthrough obtained in this approach is a precise account of the Bekenstein-Hawking entropy formula and its generalizations in terms of a microscopic counting for several BPS configurations, see~\cite{Sen:2014aja} for a recent review. It is very interesting to go beyond the counting problem and ask if the detailed understanding of the microstates of supersymmetric black holes can be used to shed any light on the conceptual puzzles that arise when formulating quantum mechanics in a black hole background. Of course this motivation underlines many developments, including the fuzzball proposal~\cite{Mathur:2005zp,Mathur:2008nj} which aims to use string theory to detect deviation from the standard general relativity picture of a black hole at the scales of the horizon. 

Here we use the AdS/CFT duality as a tool to study a particularly simple set of heavy operators $O_H$ in D1D5 CFT which are the Ramond-Ramond (RR) ground states. This ensemble is not dual to a macroscopic black hole at the level of two derivative gravity\footnote{See \cite{Sen:2009bm} for a critical discussion of this system.}, but it provides a good testing ground as we know in detail the gravitational solutions dual to these states~\cite{Lunin:2001jy,Lunin:2002iz,Kanitscheider:2007wq}. It is possible to test the dictionary between the RR ground states on the CFT side and the corresponding bulk description in terms of smooth geometries~\cite{Skenderis:2006ah,Kanitscheider:2006zf,Kanitscheider:2007wq,Taylor:2007hs,Giusto:2015dfa}: the basic idea is to exploit the AdS/CFT map between protected CFT operators $O_L$ and the supergravity modes in the bulk and then compare the 3-point CFT correlators $\langle O_H  O_H O_L\rangle$ with the holographic results obtained from the dual microstate geometries. Here the supergravity operators are indicated with a subscript $L$ because they are ``light'', meaning that their conformal dimension is fixed in the large central charge limit $c= 6 N \to\infty$. This class of 3-point correlators is protected~\cite{Baggio:2012rr} and so it is possible to match directly the results obtained in the weakly curved gravitational regime and those derived at a different point in the D1D5 SCFT moduli space, where the boundary theory can be described in terms of a free orbifold. 

While focusing on non-renormalized quantities is useful to established a dictionary between BPS states in different descriptions, this type of observables is not best suited to study interesting gravitational features of the black hole microstates. So it is important to extend the analysis to non-protected quantities involving heavy operators. Two dynamical quantities of this type have been under detailed scrutiny: the entanglement entropy of a region in a non-trivial state~\cite{Bhattacharya:2012mi,Giusto:2014aba,Asplund:2014coa} and the HHLL 4-point function with two heavy and two light operators
\begin{equation}
  \label{eq:4corri}
  \langle O_H(z_1, \bar{z}_1) \bar{O}_H (z_2,\bar{z}_2) O_L( z_3, \bar{z}_3) \bar{O}_L(z_4,\bar{z}_4)\rangle~.
\end{equation}
In this paper we study this second observable\footnote{There is a vast literature on holographic four point correlators in the context of the AdS$_5 /{\cal N}=4$ SYM duality, see~\cite{Rastelli:2017udc} for a detailed discussion of a modern approach to the problem and references to original papers. Here we focus on the AdS$_3$/CFT$_2$ case and the HHLL correlators on which much less is known.} focusing on the large central charge limit $c\gg 1$. When the D1D5 SCFT is at the free orbifold point in its moduli space, it is possible to calculate the correlator~\eqref{eq:4corri} exactly by using standard techniques and to study the statistical properties of the result when the heavy operator is chosen from an ensemble of RR ground states~\cite{Balasubramanian:2005qu,Balasubramanian:2016ids}. In order to extract detailed information on the dual gravitational theory, it is of course important also to deform the SCFT away from the free orbifold point and a possible avenue for doing this is to insert perturbatively operators corresponding to the interesting superconformal deformations (see~\cite{Carson:2014ena} and reference therein for a recent discussion of this approach). Here we focus on the opposite limit and discuss how to calculate~\eqref{eq:4corri} directly in the strongly interacting regime where the SCFT is well approximated by type IIB supergravity. 

Notice that it is not straightforward to use the technology of the Witten diagrams to calculate the correlators above, since the heavy states correspond to multi-particle operators with a large conformal dimension and are not dual to a single supergravity mode. We bypass this issue by exploiting the known smooth geometries dual to the heavy states; then we use the standard AdS/CFT dictionary to calculate the HHLL correlators by studying the quadratic fluctuations of the supergravity field dual to the light operators in the asymptotycally AdS geometry describing the heavy operators. This technique was developed~\cite{Galliani:2016cai,Galliani:2017jlg} in several concrete examples in the AdS$_3$/CFT$_2$ context which is of interest for this paper. In particular, these works discussed the case where the light operator is a simple chiral primary operator (see~\eqref{eq:Ofer}): \cite{Galliani:2016cai} focused on the case where the heavy state is made out of many copies of the same supergravity mode and found that the 4-point correlator at the gravity point matched precisely the orbifold theory result, suggesting that there is a non-renormalization theorem for this type of correlators; \cite{Galliani:2017jlg} considered a more complicated heavy operator made out of two types of supergravity modes. This second case provides the first explicit example of a dynamical HHLL correlator, where the result in the SCFT strong coupling region is radically different from the one valid at the orbifold point. However, the quadratic equations around the asymptotically AdS geometry were explicitly solved in a particular approximation where the two constituents forming the heavy multi-particle state are not on the same footing: the modes carrying a non-trivial R-charge are much more numerous than the modes with no R-charge. In this limit, the HHLL correlators could be written in terms of the standard D-functions that appear also in the evaluation of the standard Witten diagrams.

In this work we generalize the analysis of~\cite{Galliani:2017jlg} in several directions. First we consider the bosonic light operator studied in~\cite{Balasubramanian:2005qu,Balasubramanian:2016ids} (see~\eqref{eq:Obos}) which is a superdescendant of the chiral primary operator mentioned above. This implies that the HHLL correlators derived in this paper should satisfy a Ward identity linking them to the correlators computed in ~\cite{Galliani:2017jlg} (see~\eqref{eq:WIg}); as a consistency check, when we specify our new supergravity results to the heavy state considered in~\cite{Galliani:2017jlg}, we show that the Ward identity is satisfied. On the gravity side, the derivation of the HHLL correlators is drastically simplified with respect to~\cite{Galliani:2017jlg} because the gravity perturbation dual to the light operator is described by the scalar Laplace equation in six dimension, while for the case of the CPO one had to deal with a coupled system of a scalar and a 3-form. This simplified setup allows to consider more general heavy operators that are formed by many different types of supergravity modes. In one approach we still keep the approximation where the heavy state constituents include a large number $N^{(++)}_1$ of R-charge carrying modes, that we denote by $|\!++\rangle_1$, and much smaller numbers $N^{(0)}_k$ of different modes with no R-charge, denoted by $|00\rangle_k$, with $k$ any positive integer. These states form an ensemble, whose generic elements we represent schematically as
\begin{equation}\label{eq:ensembleD1D5}
(|\!++\rangle_1)^{N_{1}^{\left(++\right)}}\prod_{k} (|00\rangle_k)^{N_k^{(0)}}\quad\mathrm{with}\quad N_{1}^{\left(++\right)}+\sum_k k N_k^{(0)}=\frac{c}{6}\,.
\end{equation}
Of course these states have a large R-charge $J\sim N_{1}^{\left(++\right)}$, but their ensemble has interesting statistical properties~\cite{Iizuka:2005uv,Balasubramanian:2005qu} and an entropy that scales like $\sqrt{c/6-J}$. One of the results of this paper is an explicit expression for the correlator~\eqref{eq:4corri} with this type of heavy states, in the supergravity region of the SCFT moduli space. In an alternative approach we focus on a RR ground state that was considered also in~\cite{Galliani:2017jlg} and is made out of only the $|\!++\rangle_1$ and $|00\rangle_1$ modes. However we keep the ratio $N_k^{(0)}/N^{(++)}_1$ of the two constituents arbitrary and derive an expression for the HHLL in terms of a Fourier series. While we do not perform the transformation to configuration space in general, we show explicitly that, when it is possible to compare them, the results obtained in the two approaches agree. 

In summary our main results are:

(i) the holographic computation of the correlator of the two bosonic operators in~\eqref{eq:Obos} in a generic state of the ensemble \eqref{eq:ensembleD1D5} in the limit $N_k^{(0)}\ll N_{1}^{\left(++\right)}$ (see~\eqref{eq:boscorr1});

(ii) the verification that the bosonic correlator computed here is related via a supersymmetric Ward identity to the fermionic correlator of~\cite{Galliani:2017jlg};

(iii) the holographic computation of the same correlator in a state with $N_k^{(0)}=0$ for $k\ge2$, exactly in the ratio $N_1^{(0)}/N^{(++)}_1$ (see~(\ref{eq:CbosCfer}-\ref{eq:b1sumln})).

One of our main motivations for performing these computations is to contrast the correlators computed in pure states with those computed in a ``black hole" background. As we mentioned above, the ensemble of BPS two-charge states is not described by a regular black hole in classical supergravity, but by the singular geometry obtained by taking the zero temperature limit of the BTZ black hole. This geometry shares some properties with black holes: in particular, as we recall towards the end of Section~\ref{sec:exb1}, correlators computed in this background vanish at large Lorentzian time, albeit only polynomially. As first pointed out in \cite{Maldacena:2001kr}, and more recently emphasised in \cite{Fitzpatrick:2016ive} in the AdS$_3$ context, the late-time decay of correlators is one of the manifestations of the information loss problem. By contrast correlators in pure states should not decay. It is easy to see that this is the case for correlators computed at the orbifold point in a generic D1D5 state \cite{Balasubramanian:2005qu,Balasubramanian:2016ids}. The orbifold point CFT, however, has some special features that distinguish it from the point where a weakly coupled gravitational description is applicable: in particular there exists at the orbifold point an infinite series of conserved (bosonic) currents, of which only the Virasoro and the R-currents survive at a generic point. The presence of these currents can certainly change qualitatively  the late-time behaviour of the correlators. In some cases, like the ones considered in~\cite{Galliani:2016cai}, even just the R-current is sufficient to completely constrain the form of the correlator, and prevent the vanishing at late times. A mechanism based on the R-current, even if it applies uniformly on the moduli space, can reasonably be argued to be non-generic \cite{Fitzpatrick:2016mjq}. The correlator we consider in this paper, where the light operators are the non-chiral primaries in \eqref{eq:Obos}, is not constrained by the R-symmetry. This is  confirmed by the fact, that we verify in Section~\ref{sec:OPE}, that only the conformal block of the identity\footnote{As explained in Section~\ref{sec:OPE}, it is convenient to use the Virasoro blocks defined with respect to the ``reduced" Virasoro generators, given by the full Virasoro's minus their R-current Sugawara contribution.} contributes to the correlator in the light-cone OPE limit. We can use the exact strong coupling result obtained in Section~\ref{sec:exb1} to analyse the late-time structure of this correlator, and even in this more generic case we find that it does not decay. Note that this conclusion applies to a correlator computed in supergravity, and hence at leading order in the $1/N$ expansion. Since all large $N$ Virasoro blocks\footnote{For a derivation of Virasoro blocks in the limit of large central charge from AdS$_3$ gravity see \cite{Hijano:2015qja}.} vanish at late times \cite{Fitzpatrick:2015zha}, the only mechanism by which we can explain our findings is that even our non-protected correlator receives contributions from an infinite series of Virasoro primaries\footnote{The contribution of these primaries should be relevant also at finite values of the central charge, as each exact individual Virasoro block is still expected to decay at late times~\cite{Chen:2017yze}.}. These primaries cannot be single-particle operators: such operators, indeed, are either dual to protected supergravity modes, but then their contribution appears already in the orbifold-point result, or to string modes, which acquire large anomalous dimensions and decouple when one moves towards the supergravity regime. So the Virasoro primaries that contribute to our correlator at strong coupling must be multi-particle operators. It would be interesting to characterize more in detail such primaries and study their anomalous dimensions and three-point functions which, as in the AdS$_5$/CFT$_4$ case, are expected to receive corrections of order $1/N$ in a generic point of the moduli space. 

We conclude this introduction by outlining the structure of the paper. We begin in Section~\ref{sec:orbCFT} by defining the ingredients of the correlators we consider and by recalling their computation at the orbifold point in the CFT moduli space. Section ~\ref{sec:bulkc} contains the holographic derivation of the correlators, which follows from the solution of the Laplace equation in the geometries dual to the RR ground states \eqref{eq:ensembleD1D5}. We first perform the perturbative computation in the limit $N_k^{(0)}\ll N_{1}^{\left(++\right)}$ and then, for a particular state, the exact computation in $N_1^{(0)}/N^{(++)}_1$. To clarify the CFT meaning of the holographic result, we take in Section~\ref{sec:OPE} various OPE limits of the strong coupling correlator: we show that that in the light-cone OPE limit the only contribution comes from the Virasoro block of the identity, but the usual Euclidean OPE contains an infinite series of Virasoro primaries. In Section~\ref{sec:ltb} we examine the late time behaviour of the correlator and find a qualitative difference with the zero-temperature limit of the thermal correlator. We summarise our results and present possible future developments in Section~\ref{sec:outlook}. Some orbifold CFT technology is reviewed in Appendix~\ref{sec:AppB}. In Appendix~\ref{sec:appA} we show that the linearised equation of motion describing our light operators reduces to the Laplace equation in six-dimensions. Some of the computational details of the holographic derivation of the correlators are explained in Appendix~\ref{sec:AppC}.

%


\section{Correlators with RR ground states}
\label{sec:orbCFT}

In this section we use the D1D5 CFT at the orbifold point to describe the correlators under analysis. In this case the CFT target space is $({\cal M}_4)^N/S_N$ (where ${\cal M}_4$ can be $T^4$ or $K_3$) and the theory can be formulated in terms of $N$ groups of free bosonic and fermionic fields\footnote{We summarise the definitions and the basic properties of the orbifold D1D5 CFT in appendix~\ref{sec:AppB}.}
\begin{equation}
  \label{eq:ff}
  \Big(\partial X^{A\dot{A}}_{(r)}(z),\, \psi^{\alpha \dot{A}}_{(r)} (z)\Big)~,~~~
  \Big(\bar\partial X^{A\dot{A}}_{(r)}(\bar z),\, \tilde\psi^{\dot{\alpha} \dot{A}}_{(r)} (\bar z)\Big)~,
\end{equation}
where $(A\dot{A})$ is a pair of $SU(2)$ indices forming a vector in the CFT target space, while $(\alpha,\dot{\alpha})$ are indices of $SU(2)_L\times SU(2)_R$ which is part of the R-symmetry group; finally $r=1,\ldots N$ is a flavour index running on the various copies of the target space on which the symmetric group $S_N$ acts. As standard in orbifold constructions, beside the untwisted sector where the fields on each copy are periodic, there are twisted sectors (labelled by conjugacies classes of $S_N$) where a group of $k$ copies form a ``strand'' and the periodicities act non-diagonally on the index $(r)$, as for instance in~\eqref{eq:bck}.

As mentioned in the introduction we study the 4-point functions with two primary light operators that are part of a short supersymmetric multiplet and two heavy operators that are RR ground states. The most general heavy state in this sector is defined by a partition of $S_N$ determining the strand structure and by the quantum numbers under the $SU(2)$'s mentioned above determining the fermionic vacuum of each strand. We focus on the ``elastic'' case, where the OPE between the two light operators and the one between the two heavy operators contain the identity and so we we have $h_H=\bar{h}_H=c/24$ and $h_L=\bar{h}_L$. Then projective invariance implies
\begin{equation}
  \label{eq:4corr}
  \langle O_H(z_1, \bar{z}_1) \bar{O}_H (z_2,\bar{z}_2) O_L( z_3, \bar{z}_3) \bar{O}_L(z_4,\bar{z}_4)\rangle = \frac{1}{z_{12}^{2h_H} z_{34}^{2 h_L}} \frac{1}{\bar{z}_{12}^{2\bar{h}_H} \bar{z}_{34}^{2 \bar{h}_L}} { \cal G}(z, \bar{z})\;,
\end{equation}
where ${\cal G}$ is a function of the projective invariant cross-ratio
\begin{equation}
  \label{eq:zcr}
  z = \frac{z_{14} z_{23}}{z_{13}z_{24}}~,~~~~
  \bar{z} = \frac{\bar{z}_{14} \bar{z}_{23}}{\bar{z}_{13}\bar{z}_{24}}
\end{equation}
and $z_{ij}=z_i-z_j$. In order to easily isolate ${\cal G}$ from the correlators one can take the gauge $z_2\to\infty$, $z_1= 0$ and $z_3=1$, which implies $z=z_4$:
\begin{equation}
  \label{eq:CcalG}
  \langle  \bar{O}_H |  O_L(1) \bar{O}_L(z,\bar{z}) |O_H\rangle \equiv {\cal C}(z,\bar{z}) =  \frac{1}{(1-z)^{2 h_L}} \frac{1}{(1-\bar{z})^{2 \bar{h}_L}} { \cal G}(z, \bar{z})\;.
\end{equation}

This type of correlators was first discussed at the orbifold point in~\cite{Balasubramanian:2005qu} where the light states were identified with one of the sixteen untwisted marginal operators corresponding to the deformations of the $T^4$. For the sake of concreteness we can choose
\begin{equation}
  \label{eq:Obos}
  O_L \to O_{\rm bos} = \sum_{r=1}^N \frac{\epsilon_{\dot{A}\dot{B}}}{\sqrt{2 N}} \partial X_{(r)}^{1\dot{A}} \bar{\partial} X_{(r)}^{1\dot{B}}~,~~~
  \bar{O}_L \to \bar{O}_{\rm bos} = \sum_{r=1}^N \frac{\epsilon_{\dot{A}\dot{B}}}{\sqrt{2 N}} \partial X_{(r)}^{2\dot{A}} \bar{\partial} X_{(r)}^{2\dot{B}} ~.
\end{equation}
With the above choice of light and heavy operators the correlator at the orbifold point depends only on the strand structure, but not on the particular quantum numbers of the RR ground state considered (this simply because the elementary bosonic and fermionic fields in~\eqref{eq:ff} commute). A standard way to calculate this correlator is to diagonalize the boundary conditions (as summarized in appendix~\ref{sec:AppB}) and then to take the linear combination of the contributions of each strand~\eqref{eq:Gbosw}
\begin{align}
  \label{eq:corrbos}
  \mathcal{C}^{\rm bos}  
= \frac{1}{N} \sum_{k=1}^N N_k \mathcal{C}_k^{\rm bos} 
=  \frac{1}{N} \sum_{k=1}^N N_k \partial \bar\partial\left[\frac{1- z \bar{z}}{(1-z)(1-\bar{z})\left(1-(z \bar{z})^\frac{1}{k} \right)} \right] \,,
\end{align}
where $N_k$ here is the number of strands of length or winding $k$ (regardless of their particular RR ground state) and we used~\eqref{eq:Gbosz}. We can express the result in terms of the cylinder coordinates $w$ ($z = e^{-\ii w}$ and $\bar{z} = e^{\ii \bar{w}}$) by using~\eqref{eq:Gbosw} for $\mathcal{C}_k^{\rm bos}$, and in this case we obtain Eq.~(4.11) of~\cite{Balasubramanian:2005qu}.

It is interesting to compare this result with the correlators where the light operator $O_L$ is the following chiral primary~\cite{Balasubramanian:2005qu,Galliani:2016cai,Galliani:2017jlg}
\begin{equation}
  \label{eq:Ofer}
  O_L \to O_{\rm fer} = \sum_{r=1}^N \frac{-i \epsilon_{\dot{A} \dot{B}}}{\sqrt{2 N}} \psi_{(r)}^{1 \dot{A}}\, \tilde{\psi}_{(r)}^{\dot{1} \dot{B}}~,~~~
  \bar{O}_L \to \bar{O}_{\rm fer} = \sum_{r=1}^N \frac{-i \epsilon_{\dot{A} \dot{B}}}{\sqrt{2 N}} \psi_{(r)}^{2 \dot{A}}\, \tilde{\psi}_{(r)}^{\dot{2} \dot{B}}~.
\end{equation}
It is again straightforward to calculate the correlator at the orbifold point by diagonalizing the boundary conditions of the fermions, see appendix~A of~\cite{Galliani:2016cai} for our conventions. However in this case the result depends on the particular RR ground state of each strand. The contribution of a strand of length $k$ and $SU(2)_L\times SU(2)_R$ quantum numbers $j=\bar{j}=1/2$ is
\begin{equation}
  \label{eq:c1212}
\mathcal{C}_{k\,\left(\frac{1}{2} \frac{1}{2}\right)}^{\rm fer} = \frac{1}{|z|} \frac{1- z \bar{z}}{(1-z)(1-\bar{z})\left(1-(z \bar{z})^\frac{1}{k} \right)} \;.
\end{equation}
The contribution from strands with general R-charge quantum numbers is given by~\eqref{eq:fkj} and the generic correlator with fermionic light operators is
\begin{equation}
  \label{eq:corrfer}
\mathcal{C}^{\rm fer} = \frac{1}{N} \sum_{k=1}^N \sum_{s=1}^8 N_{k}^{\left(s\right)}
\mathcal{C}_{k\,\left(s\right)}^{\rm fer}\;,
\end{equation}
where $\mathcal{C}_{k\,\left(s\right)}^{\rm fer}$ is defined in~\eqref{eq:Cjbj}, $s$ runs over the 8 different RR ground states (4 with $j,\bar{j}=\pm 1/2$ and 4 with $j,\bar{j}=0$), $N_{k}^{\left(s\right)}$ is the number of strands of length $k$ in the state $s$, which has to satisfy the constraint $ \sum_{s,\,k} k N_{k}^{\left(s\right)} = N$. It is convenient to indicate each strand as ket-vectors displaying its $j,\bar{j}$ quantum numbers and its winding $k$
\begin{equation}
  \label{eq:ketdef}
  | \pm\pm\rangle_k\;, ~~|00 \rangle_k^{(\dot{A} \dot{B})}~~ \mbox{and}~~ |00\rangle_k\;. 
\end{equation}
The last type of strand is a scalar of all $SU(2)$ mentioned at the beginning of this section and will play a particular role in the heavy states we consider in our supergravity analysis. Then a general RR ground state is just an arbitrary tensor product of the ket-vectors in~\eqref{eq:ketdef} provided that the total winding is $N$. Notice that, despite the fact that the fermionic correlator is sensitive to the $SU(2)$ quantum numbers of each strand, the $\partial$ and $\bar\partial$ derivative of $|z| \mathcal{C}_{k\,(j\,\bar{j})}^{\rm fer}$ is independent of $j$, $\bar j$ and matches the structure in~\eqref{eq:corrbos}. Thus we have $\mathcal{C}^{\rm bos} = \partial \bar\partial\left(|z| \mathcal{C}^{\rm fer}\right)$ when the heavy state is an arbitrary RR ground state. We will now show that this is a consequence of a simple Ward identity.

The bosonic operator $O_{\rm bos}$ in~\eqref{eq:Obos} is a superdescendant of the chiral primary $O_{\rm fer}$ in~\eqref{eq:Ofer}. At the orbifold point this can be easily checked by using 
\begin{equation}
  \label{eq:wO}
\begin{aligned}
\oint\limits_{w\sim z} \frac{dw}{2\pi\ii}\,\sqrt{w}\,G^1_1(w)\,\psi^{2\dot{C}}(z) &= \sqrt{z}\,\partial X_{1\dot{E}}(z)\,\epsilon^{\dot{E}\dot{C}} 
\\
  \oint\limits_{w\sim z} \frac{dw}{2\pi\ii}\,\sqrt{w}\,G^{\alpha}_A(w)\,\partial X^{B\dot{B}}(z) &= \delta_{A}^B \left( \sqrt{z}\,\partial \psi^{\alpha\dot{B}}(z) + \frac{\psi^{\alpha\dot{B}}}{2\sqrt{z}} \right) =  \delta_A^B\,\partial_z \left( \sqrt{z}\,\psi^{\alpha\dot{B}}(z) \right)\;,
\end{aligned}
\end{equation}
which follow from the OPE contractions summarised in appendix~\ref{sec:AppB}, with similar equations holding in the anti-holomorphic sector. As usual, we can start for instance from the bosonic correlator and write one operator, for example that in $z=1$, in terms of the supersymmetry variation in the first line of~\eqref{eq:wO}; we then deform the contour of integration so that it goes around all the other insertions in the correlator~\eqref{eq:CcalG}. This explains why in~\eqref{eq:wO} we inserted an extra factor of $\sqrt{w}$ which makes the integration of the supercurrents around the RR states at $z=0,\infty$ well defined. Since we are focusing on the case where $O_H$ are RR ground states, the contributions from $w\sim 0$ and $w\sim \infty$ vanish and so the only non trivial terms come from $w\sim z$ and $\bar{w}\sim \bar{z}$, which can be computed using the second line of~\eqref{eq:wO}. In summary we obtain the relation mentioned above
\begin{equation}
  \label{eq:WIg}
  \langle  \bar{O}_H |  O_{\rm bos} (1) \bar{O}_{\rm bos} (z,\bar{z}) |O_H\rangle = \partial \bar\partial \Big[ |z| \langle  \bar{O}_H |  O_{\rm fer}(1) \bar{O}_{\rm fer} (z,\bar{z}) |O_H\rangle \Big]\;.
\end{equation}
This is clearly satisfied by the orbifold point results~\eqref{eq:corrbos} and~\eqref{eq:corrfer}, but since this relation uses only the superconformal algebra, it holds at a generic point of the CFT moduli space and in the next section we will check its validity in the supergravity limit.


\section{Bosonic correlators at strong coupling}
\label{sec:bulkc}

The aim of this section is to study the HHLL correlators discussed above on the bulk side by using the supergravity approximation of type IIB string theory on AdS$_3 \times S^3 \times {\cal M}$. The case where the light operators are the chiral primaries~\eqref{eq:Ofer} was discussed in~\cite{Galliani:2017jlg}, so here we consider the correlators with the bosonic light operators of dimension two given in \eqref{eq:Obos}. While in the orbifold CFT description it was easy to keep the RR ground states completely generic, in the bulk analysis we will find it convenient to focus on a subsector of these heavy states. First we focus on the states that are invariant under the $SU(2)$'s acting on the coordinates of ${\cal M}_4$, which ensures that the dual solutions are invariant under rotations of the four stringy-sized compact directions. Then we focus on the case where the RR ground states are made of a large number $N_{1}^{\left(++\right)}$ of strands of the type $|\!++\rangle_1$ (of winding one and $j=\bar{j}=1/2$) while the remaining strands have arbitrary winding $k\ge 1$ but are in the unique RR state $s=0$ that is a scalar of all $SU(2)$'s; we denote strands of this type as $|00\rangle_k$ and their numbers as $N_k^{(0)}$. These states form the ensemble that was introduced in~\eqref{eq:ensembleD1D5}. On the bulk side the restriction to this subset of states simplifies the 6D metric~\eqref{eq:6DE}. The family of D1D5 geometries dual to these states has in fact played an important role in some recent supergravity  developments \cite{Bena:2015bea,Bena:2016agb,Bena:2016ypk}. At some point of our analysis we will also assume that the numbers of $|00\rangle_k$ strands are parametrically smaller than the number of $|\!++\rangle_1$ strands ($N_k^{(0)}\ll N_{1}^{\left(++\right)}$): this will allow the perturbative approach in $b_k$ discussed in Section~\ref{sec:perturbative}.

The heavy operators $O_H$ are described in the gravity regime by 6D geometries that asymptotically approximate AdS$_3\times S^3$ and are everywhere regular and horizonless. Operators that are Ramond ground states both in the left and in the right sector are dual to geometries carrying D1 and D5 charges but no momentum charge. The six-dimensional Einstein metric dual to RR ground states that are invariant under rotations in the four compact dimensions is~\cite{Lunin:2001jy,Lunin:2002iz,Kanitscheider:2007wq}
\begin{equation}\label{eq:6DE}
ds^2_6=-\frac{2}{\sqrt{\mathcal{P}}}(dv+\beta)(du+\omega)+\sqrt{\mathcal{P}}\,ds^2_4\,,
\end{equation}
with
\begin{equation}\label{eq:calP}
\mathcal{P}\equiv Z_1 Z_2 - Z_4^2\,.
\end{equation}
We use light-cone coordinates
\begin{equation}
u\equiv\frac{t-y}{\sqrt{2}}\,,\quad v\equiv\frac{t+y}{\sqrt{2}}\,,
\end{equation}
with $t$ time and $y$ the coordinate along $S^1$, and denote by $ds^2_4$ the flat metric on $\mathbb{R}^4$. $Z_1$, $Z_2$, $Z_4$ are harmonic scalar functions on $\mathbb{R}^4$ and $\beta$, $\omega$ are one-forms with self-dual and anti-self-dual 2-form field strengths. Apart from the metric, all other fields of type IIB supergravity are non-trivial in the solution: their expressions are given in \eqref{eq:otherfields}, but will not be relevant for the correlator we compute here. 

The form of the supergravity data $Z_1$, $Z_2$, $Z_4$, $\beta$ and $\omega$ depends on the RR ground state and is generically complicated. As mentioned above, we focus on the family of D1D5 states described in \eqref{eq:ensembleD1D5}. The dual gravity solutions depend on some continuous parameters: $a$, whose square is proportional to $N_{1}^{\left(++\right)}$, and $b_k$, whose square is proportional to $k N_k^{(0)}$ \cite{Giusto:2015dfa}:
\begin{equation}
N_{1}^{\left(++\right)}=N\,\frac{a^2}{a_0^2}\,,\quad k N_k^{(0)}=N\,\frac{b_k^2}{2 a_0^2}\quad\mathrm{with}\quad a_0^2 \equiv \frac{Q_1 Q_5}{R^2}\,.
\end{equation}
Here $R$ is the radius of the CFT circle and $Q_1$, $Q_5$ are the supergravity D1 and D5 charges, related to the numbers $n_1$, $n_5$ of D1 and D5 branes by
\begin{equation}
Q_1 = \frac{(2\pi)^4\, n_1\, g_s\, \alpha'^4}{V_4}\,,\quad Q_5 = n_5\, g_s \,\alpha'\,, 
\end{equation}
with $g_s$ the string coupling and $V_4$ the volume of $T^4$.
The condition that the total number of strands be $N$ implies the constraint
\begin{equation}\label{eq:constraint}
a^2 + \sum_k \frac{b_k^2}{2}=a_0^2\,,
\end{equation}
which turns out to be also the regularity condition for the metric. The metrics are more easily written in spheroidal coordinates in which the flat $\mathbb{R}^4$ metric is
\begin{equation}\label{eq:ds4}
ds^2_4 = \Sigma \left(\frac{dr^2}{r^2+a^2}+d\theta^2\right)+(r^2+a^2)\sin^2\theta d\phi^2+r^2\cos^2\theta d\psi^2\,,\quad \Sigma\equiv r^2+a^2 \cos^2\theta\,.
\end{equation}
The remaining data encoding the metric are
\begin{subequations}\label{eq:solb}
\begin{equation}
\begin{aligned}
Z_1 = &\, \frac{R^2}{Q_5\,\Sigma}\,\Biggl[a_0^2+ \sum_{k,k'} \frac{b_k b_{k'}}{2}\,\frac{a^{k+k'}}{(r^2+a^2)^{\frac{k+k'}{2}}}\,\sin^{k+k'}\theta \cos((k+k')\phi) \\
&\,+\sum_{k>k'} b_k b_{k'}\,\frac{a^{k-k'}}{(r^2+a^2)^{\frac{k-k'}{2}}}\,\sin^{k-k'}\theta \cos((k-k')\phi)\Biggr]\,,\quad Z_2 = \frac{Q_5}{\Sigma}\,,\\
\end{aligned}
\end{equation}
\begin{equation}
Z_4 = \frac{R}{\Sigma}\,\sum_k b_k\frac{a^{k}}{(r^2+a^2)^{\frac{k}{2}}}\,\sin^{k}\theta \cos(k\phi)\,,\\
\end{equation}
\begin{equation}
\beta=\frac{R\,a^2}{\sqrt{2}\,\Sigma}\,(\sin^2\theta d\phi-\cos^2\theta d\psi)\,,\quad \omega=\frac{R\,a^2}{\sqrt{2}\,\Sigma}\,(\sin^2\theta d\phi+\cos^2\theta d\psi)\,.
\end{equation}
\end{subequations}
For generic values of $b_k$ the geometry is complicated, but it can be shown to be regular and without horizon for any values of the parameters, as far as the constraint \eqref{eq:constraint} is satisfied.

\subsection{The perturbation}
To compute the correlator of two light and two heavy operators one should consider the wave equation for a perturbation in the background \eqref{eq:6DE}. 
The bosonic light operator $O_L=O_\mathrm{bos}$ is described by a minimally coupled scalar in the 6D Einstein metric $ds^2_6$. We show in Appendix~\ref{sec:appA} that such scalars arise by dimensional reduction from traceless perturbations of the metric on $T^4$, and thus have the right quantum numbers to be dual to the CFT operators $\partial X^{(i} \bar \partial X^{j)}$, with $i,j=1,\ldots,4$.

Following the logic of \cite{Galliani:2016cai,Galliani:2017jlg}, the gravity computation of the correlator requires solving the wave equation
\begin{equation}\label{eq:box6}
\square_6 B=0\,,
\end{equation}
where $\square_6$ is the scalar Laplace operator with respect to $ds^2_6$
\begin{equation}\label{eq:lapl6}
\square_6 \cdot \equiv \frac{1}{\sqrt{g_6}} \,\partial_M (\sqrt{g_6} \,g^{M N}_6 \partial_N \cdot)\,,
\end{equation}
with the boundary condition
\begin{equation}\label{eq:boundary}
B \sim \delta(t,y) + \frac{b(t,y)}{r^2}
\end{equation}
for large $r$. Since the background metric is regular everywhere, one should also require that $B$ have no singularities at any finite value of $r$. As the operator $O_L$ is an R-charge singlet, only the projection of $B$ on the trivial scalar spherical harmonic on $S^3$ contributes to our correlator. The 4-point function computed on the Euclidean plane is encoded in the function $b(t,y)$ via
\begin{equation}\label{eq:4ptb}
\langle O_H(0) \bar O_H(\infty) O_L(1,1) \bar O_L(z,\bar z)\rangle =\frac{1}{|1-z|^4}\mathcal{G}^\mathrm{bos}(z,\bar z)= (z \bar z)^{-1} \,b(z,\bar z)\,,
\end{equation}
where
\begin{equation}
z= e^{i\frac{t+y}{R}}=e^{\frac{t_e+i y}{R}}\,,\quad \bar z= e^{i\frac{t-y}{R}}=e^{\frac{t_e-i y}{R}}\,,
\end{equation}
with $t_e\equiv i t$ the Euclidean time. The factor $(z \bar z)^{-1}$ on the r.h.s. of \eqref{eq:4ptb} comes from the transformation of the primary field $\bar O_L(z,\bar z) = (z \bar z)^{-1}\,\bar O_L(t,y)$ from the cylinder to the plane coordinates.   

The laplacian in \eqref{eq:lapl6} is most easily derived if one writes the 6D metric as if one were performing a dimensional reduction on $S^3$ \cite{Giusto:2014aba,Giusto:2015dfa,Bena:2017upb}:
\begin{equation}
ds^2_6 = V^{-2} g_{\mu\nu} dx^\mu dx^\nu + G_{\alpha\beta} (dx^\alpha + A^\alpha_\mu dx^\mu)(dx^\beta+A^\beta_\nu dx^\nu)\,,
\end{equation}
where
\begin{equation}
V^2\equiv \frac{\mathrm{det} \,G}{(Q_1 Q_5)^{3/2} \sin^2\theta \cos^2\theta}\,. 
\end{equation}
We have split the 6D coordinates in the AdS$_3$ coordinates $x^\mu, x^\nu,\ldots\equiv (r,t,y)$ and the $S^3$ coordinates $x^\alpha, x^\beta,\ldots\equiv (\theta,\phi,\psi)$. The definition of $g_{\mu\nu}$, $G_{\alpha\beta}$, $A^\alpha_\mu$ depends of course on the choice of coordinates: the coordinates are fixed at the boundary by the requirement that the metric looks like AdS$_3\times S^3$ asymptotically, but one is free to redefine the coordinates in the space-time interior. For lack of a better choice, we will stick to the coordinates defined in \eqref{eq:ds4}.

If one takes the solution in \eqref{eq:solb} and sets $b_k=0$ for any $k$, one finds that $g_{\mu\nu}$ becomes the metric of global AdS$_3$
\begin{equation}
g_{\mu\nu} dx^\mu dx^\nu \Bigl |_{b_k=0} = \sqrt{Q_1 Q_5} \left[\frac{dr^2}{r^2+a_0^2}-\frac{r^2+a_0^2}{Q_1 Q_5}dt^2 + \frac{r^2}{Q_1 Q_5} dy^2\right]\equiv  \sqrt{Q_1 Q_5} \,ds^2_{AdS_3}
\end{equation}
and $G_{\alpha\beta}$ the metric of the round $S^3$. When, like in this case, the metric $g_{\mu\nu}$ does not depend on the coordinates of $S^3$, the 6D Laplace equation \eqref{eq:box6} admits an $S^3$-independent solution which satisfies the simpler equation
\begin{equation}\label{eq:box3}
\square_3 B=0\,,
\end{equation}
with $\square_3$ the laplacian of $g_{\mu\nu}$:
\begin{equation}
\label{eq:Box3}
\square_3 \cdot \equiv \frac{1}{\sqrt{g}} \,\partial_\mu (\sqrt{g} \,g^{\mu\nu} \partial_\nu \cdot)\,.
\end{equation}

In general however the 6D metric does not factorise and $g_{\mu\nu}$ and $G_{\alpha\beta}$ depend on both AdS$_3$ and $S^3$ coordinates. In this situation solving the 6D equation \eqref{eq:box6} exactly seems hard. When this happens one can resort to an approximation scheme that was used already in \cite{Galliani:2017jlg}: we solve the wave equation perturbatively in $b_k$, keeping only the first non-trivial order $\mathcal{O}(b_k^2)$. In the following we will apply this perturbative method to compute the correlator for generic $b_k$'s. In the particular example in which $b_1$ is the only non-vanishing mode, we will be able to do better and perform the computation exactly in $b_1$.

\subsection{Perturbative computation for generic  $b_k$'s}
\label{sec:perturbative}
   
We consider here a generic state in the ensemble~\eqref{eq:ensembleD1D5} and compute the correlator in the limit $N_k^{(0)}\ll N_1^{(++)}$, keeping the first non-trivial term in an expansion in $b_k/a_0$. This contribution already depends on the CFT moduli and hence it contains non-trivial dynamical informations. We perform the $b_k$-expansion keeping $Q_1$, $Q_5$ and $R$ (and hence $a_0$) fixed: on the CFT side this means we are not varying the central charge nor the size of the circle on which the CFT is defined. At zero-th order in $b_k$ the metric is AdS$_3\times S^3$, and we will expand the terms of order $b_k^2$ in the basis of spherical harmonics of this unperturbed $S^3$. We thus write the solution of \eqref{eq:box6} as
\begin{equation}
B= B_0 + B_1 + \mathcal{O}(b_k^4)\,,
\end{equation}
where $B_1$ quadratic in $b_k$. The terms of order zero and two of the wave equation give
\begin{equation}\label{eq:wavepert}
\square_0 B_0 = 0\,,\quad \square_0 B_1  = -\square_1 B_0\,,
\end{equation}
where $\square_0$ is the laplacian of global AdS$_3$ 
\begin{equation}
\square_0 \cdot  \equiv \frac{1}{r}\partial_r(r(r^2+a_0^2) \partial_r \cdot)-\frac{a_0^2 \,R^2}{r^2+a_0^2}\,\partial_t^2 \cdot +\frac{a_0^2 \,R^2}{r^2}\, \partial_y^2 \cdot\,,
\end{equation}
and $\square_1$ is the order $b_k^2$ contribution to the laplacian $\square_3$ defined in \eqref{eq:box3}. The first equation in \eqref{eq:wavepert}, together with the asymptotic boundary condition \eqref{eq:boundary} and the regularity condition, implies that $B_0$ is the usual bulk-to-boundary propagator of dimension $\Delta=2$ in global AdS$_3$:
\begin{equation}\label{eq:B0}
B_0(r,t,y) = K_2^{\mathrm{Glob}}(r,t,y|t'=0,y'=0)=\left[\frac{1}{2}\,\frac{a_0}{\sqrt{r^2+a_0^2} \,\cos (t/R) - r\,\cos (y/R)}\right]^2\,.
\end{equation}

The second equation in \eqref{eq:wavepert} is an equation for $B_1$. If the metric $g_{\mu\nu}$ is a non-trivial function on $S^3$, the $B_1$ that solves this equation has components along non-trivial $S^3$ spherical harmonics, which we should project away for the purpose of extracting the bosonic correlator. In particular all terms in the solution \eqref{eq:solb} that are proportional to $b_k b_{k'}$ for $k\not=k'$ depend non-trivially on $\phi$ as $\cos((k-k')\phi)$ and source non-trivial spherical harmonics in $B_1$: hence they do not contribute to the correlator at quadratic order in $b_k$. We can thus simplify the computation by focusing on a single $k$-mode at a time. The metric $g_{\mu\nu}$ derived from the solution where a single $b_k$ is non-vanishing is
\begin{subequations}\label{eq:g}
\begin{equation}
\frac{g^{(k)}_{tt}}{\sqrt{Q_1 Q_5}}=-\frac{r^2+a^2}{R^2 a_0^4}\left(a^2 + \frac{b_k^2}{2}\frac{r^2}{\Sigma} F_k\right)\,,\quad \frac{g^{(k)}_{yy}}{\sqrt{Q_1 Q_5}}=\frac{r^2}{R^2 a_0^4}\left(a^2 + \frac{b_k^2}{2}\frac{r^2+a^2}{\Sigma} F_k\right)\,,
\end{equation}
\begin{equation}
\frac{g^{(k)}_{rr}}{\sqrt{Q_1 Q_5}}=\frac{1}{a_0^4 (r^2+a^2)}\left(a^2 + \frac{b_k^2}{2}\frac{r^2}{\Sigma} F_k\right) \left(a^2 + \frac{b_k^2}{2}\frac{r^2+a^2}{\Sigma} F_k\right)\,,
\end{equation}
\end{subequations}
with
\begin{equation}\label{eq:Fk}
F_k \equiv 1- \left(\frac{a^2 \sin^2\theta}{r^2+a^2} \right)^k\,.
\end{equation}
We see that, unless $k=1$, even for a single mode $g_{\mu\nu}$ depends non-trivially on the $S^3$ coordinate $\theta$. To compute $B_1$, one should expand the laplacian of $g^{(k)}_{\mu\nu}$ up to order $b_k^2$ ($\square^{(k)}=\square_0 + b_k^2 \,\square_1^{(k)} + \mathcal{O}(b_k^4)$)  and project on the trivial spherical harmonic. One finds
\begin{equation}\label{eq:source}
\langle J_k \rangle \equiv -\langle \square_1^{(k)} B_0  \rangle=-\frac{r}{(r^2+a_0^2)} \partial_r B_0+\frac{a_0^2\,R^2}{(r^2+a_0^2)^2}\, \partial_t^2 B_0 +\frac{R^2}{2 a_0^2}\,S_k \,(\partial_t^2 B_0-\partial_y^2 B_0)\,,
\end{equation}
where
\begin{equation}
S_k\equiv \sum_{p=2}^k \left(\frac{a_0^2}{r^2+a_0^2}\right)^p \langle \sin^{2p-2}\theta \rangle=\sum_{p=2}^k \frac{1}{p}\left(\frac{a_0^2}{r^2+a_0^2}\right)^p \,,
\end{equation}
and the bracket $\langle \cdot \rangle$ denotes the average on $S^3$. In deriving \eqref{eq:source} we have also used that $\square_0 B_0=0$. The second equation in \eqref{eq:wavepert} is then easily integrated using the AdS$_3$ bulk-to-bulk propagator $G^{\mathrm{Glob}}_2(\mathbf{r}'|r,t,y)$, and summing over all the modes:
\begin{equation}
B_1(r,t,y) = -i\,\sum_k b_k^2 \int d^3\mathbf{r}'\, \sqrt{-g_{AdS_3}}\,G^{\mathrm{Glob}}_2(\mathbf{r}'|r,t,y)\,\langle J_k(\mathbf{r}')\rangle \,,
\end{equation}
where $\mathbf{r}'\equiv \{r',t',y'\}$ is a point in AdS$_3$ and $g_{AdS_3}$ the metric of global AdS$_3$.

According to \eqref{eq:4ptb}, the correlator is determined by the large $r$ limit of $B_1$, which follows from the asymptotic limit
of $G^{\mathrm{Glob}}_2(\mathbf{r}'|r,t,y)$: $G^{\mathrm{Glob}}_2(\mathbf{r}'|r,t,y)\to \frac{a_0^2}{2\pi r^2} K^{\mathrm{Glob}}_2(\mathbf{r}'|t,y)$. Moving from Lorentzian cylinder to Euclidean plane, one finds that the order $b^2_k$ contribution to the 4-point function is  
\begin{equation}\label{eq:integral}
\begin{aligned}
\langle O_H(0) \bar O_H(\infty) O_L(1,1) \bar O_L(z,\bar z)\rangle|_{b^2_k} = -\sum_k \frac{b^2_k}{2\pi}\int \!d^3\mathbf{w}\, \sqrt{\bar g} \,K_2(\mathbf{w}|z,\bar z)\,\langle J_k(\mathbf{w})\rangle\,,
\end{aligned}
\end{equation}
where $\bar g$ is the metric of Euclidean AdS$_3$ and $K_2(\mathbf{w}|z,\bar z)$ the usual bulk-to-boundary propagator in the Poincar\'e coordinates $\mathbf{w}$. The integral in \eqref{eq:integral}), with the source $\langle J_k \rangle$ given in \eqref{eq:source}, can be expressed in terms of D-functions using standard methods: we summarise the various steps in Appendix~\ref{sec:AppC}. Including also the free contribution at $b_k=0$, the final result for the strong coupling limit of the bosonic correlator up to order $b_k^2$ can be written in the suggestive form
\begin{equation}\label{eq:boscorr1}
\mathcal{C}^\mathrm{bos}_{{\cal O}(b^2)}(z,\bar z)=\partial {\bar \partial} \left[\frac{1}{|1-z|^2}+\sum_k \frac{b^2_k}{a_0^2}\left(-\frac{1}{2}\frac{1}{|1-z|^2}+\sum_{p=1}^k \frac{|z|^2 \hat{D}_{pp22}}{\pi\,p}\right)\right]\,.
\end{equation}
Comparing this result with the Ward identity \eqref{eq:WIg} linking bosonic and fermionic correlators, one is lead to the following natural guess for the correlator with fermionic light operators
\begin{equation}\label{eq:fercorr}
\mathcal{C}^\mathrm{fer}_{{\cal O}(b^2)}(z,\bar z)=\frac{1}{|z|}\left[\frac{1}{|1-z|^2}+\frac{b^2_1}{a_0^2} \frac{N}{2} +\sum_k \frac{b^2_k}{a_0^2}\left(-\frac{1}{2}\frac{1}{|1-z|^2}+\sum_{p=1}^k \frac{|z|^2 \hat{D}_{pp22}}{\pi\,p}\right)\right]\,.
\end{equation}
The term of order $N$ is the disconnected contribution to the correlator, which cannot be predicted by the Ward identity since it is annihilated by the operator $\partial {\bar \partial} (|z|\cdot)$.

Specialising \eqref{eq:fercorr} to the heavy state considered in \cite{Galliani:2017jlg}, which has $b_1=b\not=0$ and $b_k=0$ for $k>1$, one can verify that the above result is in perfect agreement with eq. (3.58) of \cite{Galliani:2017jlg} (thanks to eq.~(D.12a) of the same paper): this checks that the Ward identity is satisfied for this particular heavy state, and provides a quite non-trivial validation of our computations. One can also check that the bosonic correlator \eqref{eq:boscorr1} has the expected symmetry under the exchange of the points $z_3$ and $z_4$. This transformation permutes $O_L$ with $\bar O_L$ and, according to the definition \eqref{eq:Obos}, amounts to exchange the $\mathcal{M}_4$ index $A=1$ with $A=2$; since the heavy operators we consider are invariant under transformations of the compact space $\mathcal{M}_4$, the correlator should be left invariant. From the definition of $z$ \eqref{eq:zcr} one sees that the transformation $z_3\to z_4$ is equivalent to $z\to 1/z$ and thus one should have that
\begin{equation}
\mathcal{G}^\mathrm{bos}(z,\bar z) = \mathcal{G}^\mathrm{bos}(z^{-1},\bar z^{-1}) \,.
\end{equation}
That the result \eqref{eq:boscorr1} has this property follows from the symmetry of the $\hat D$-functions
\begin{equation}
\hat D_{pp22}(z^{-1},\bar z^{-1})=|z|^4 \hat D_{pp22}(z,\bar z)\,.
\end{equation}

\subsection{Exact computation for $b_k= b\, \delta_{k,1}$}
\label{sec:exb1}

The solution in which only the mode $b_1\equiv b$ is non-vanishing is particularly simple: one sees indeed from \eqref{eq:g} and \eqref{eq:Fk} that $F_1=\Sigma/(r^2+a^2)$ and thus the 3D metric $g_{\mu\nu}$ is $\theta$-independent. One can thus look for an exact solution of the 3D Laplace equation \eqref{eq:box3}:
\begin{equation}\label{eq:k=1ex}
\frac{r^2+a^2}{r(r^2+a^4/a_0^2)} \,\partial_r [ r (r^2+a^2) \partial_r B]-\frac{a_0^2}{r^2+a^4/a_0^2}\, \partial_\tau^2 B + \frac{a_0^2}{r^2}\,\partial_\sigma^2B=0\,,
\end{equation}
where we have defined
\begin{equation}
\tau\equiv \frac{t}{R}\,,\quad \sigma\equiv \frac{y}{R}\,.
\end{equation}
Our analysis here will follow the one in appendix B of \cite{Galliani:2016cai}. The solution of \eqref{eq:k=1ex} that is regular at $r=0$ and that has the asymptotic behaviour \eqref{eq:boundary} for large $r$ is
\begin{equation}
B =\frac{1}{(2\pi)^2}\sum_{l\in\mathbb{Z}}\int \!d\omega \,e^{i\omega \tau+ il\sigma}\,g(\omega,l)\,\left(\frac{r}{\sqrt{r^2+a^2}}\right)^{\!|l|} \,{}_2F_1\left(\frac{|l|+\gamma}{2},\frac{|l|-\gamma}{2},1+|l|;\frac{r^2}{r^2+a^2}\right)\,,
\end{equation}
where
\begin{equation}
g(\omega,l)=\frac{\Gamma\left(1+\frac{|l|+\gamma}{2}\right)\Gamma\left(1+\frac{|l|-\gamma}{2}\right)}{\Gamma(1+|l|)}
\end{equation}
and
\begin{equation}
\gamma\equiv \frac{\sqrt{a_0^2\, \omega^2 -\frac{1}{2}\, b^2 \,l^2}}{a}\,.
\end{equation}
The function $b(t,y)$ defined in \eqref{eq:boundary} is extracted from the large $r$ limit of $B$:
\begin{equation}
\begin{aligned}
b(\tau,\sigma)=&\,\frac{a^2}{a_0^2}\,\sum_{l\in\mathbb{Z}}\!\int \!\frac{d\omega}{(2\pi)^2}\,e^{i\omega \tau+ il\sigma} \left[-\frac{|l|}{2}+\frac{l^2-\gamma^2}{4}\left(H\left(\frac{|l|+\gamma}{2}\right)+H\left(\frac{|l|-\gamma}{2}\right)-1\right)\right] ,
\end{aligned}
\end{equation}
where $H(z)$ is the harmonic number, which is related to the digamma function $\psi(z)$ as
\begin{equation}
H(z) = \psi(z+1)+\gamma_E = \sum_{n=1}^\infty \left(\frac{1}{n}-\frac{1}{n+z}\right)\,.
\end{equation}
Discarding contact terms proportional to $\delta(\tau)$ and/or $\delta(\sigma)$ and their derivatives, and using the identity 
\begin{equation}
l^2-\gamma^2= \frac{a_0^2}{a^2}\,(l^2-\omega^2)\,,
\end{equation}
one can write
\begin{equation}
\label{eq:bdtdsbf}
b(\tau,\sigma)=\frac{\partial_\tau^2-\partial_\sigma^2}{4}\,b_F(\tau,\sigma)\,,
\end{equation}
where
\begin{equation}
b_F(\tau,\sigma)=\sum_{l\in\mathbb{Z}} \int \!\frac{d\omega}{(2\pi)^2}\,e^{i\omega \tau+ il\sigma}\,\sum_{n=1}^\infty\left(\frac{2}{\gamma-|l|-2n}-\frac{2}{\gamma+|l|+2n}\right)\,.
\end{equation}
The $\omega$-integral is performed along Feynman's contour; assuming $\tau>0$ the contour has to be closed on the upper half plane, so we pick the poles on the negative real axis:
\begin{equation}
\omega_n = - \frac{a}{a_0} \sqrt{(|l| + 2 n)^2+ \frac{b^2  l^2}{2 a^2}}\,.
\end{equation}
The correlator on the plane is found by transforming from the $(\tau,\sigma)$ coordinates to the $(z=e^{i(\tau+\sigma)},\bar z=e^{i(\tau-\sigma)})$ coordinates and using \eqref{eq:4ptb}. Dropping an irrelevant overall normalization one finds
\begin{equation}
\label{eq:CbosCfer}
\mathcal{C}^\mathrm{bos}(z,{\bar z}) = \partial {\bar \partial} \,\left (|z| \,\mathcal{C}^\mathrm{fer}(z,{\bar z})\right)\,, 
\end{equation}
with $\mathcal{C}^\mathrm{fer}(z,{\bar z}) = \mathcal{C}^\mathrm{fer}(\tau,\sigma)/|z|$, where the factor $1/|z|$ follows from the transformation of the operator in $z$, and
\begin{equation}
\label{eq:b1sumln}
\mathcal{C}^\mathrm{fer}(\tau,\sigma)=
\,\frac{a}{a_0} \sum_{l\in\mathbb{Z}} e^{il\sigma}\sum_{n=1}^\infty \frac{\exp\left[-i\frac{a}{a_0}\sqrt{(|l| + 2 n)^2+ \frac{b^2  l^2}{2 a^2}}\tau\right]}{\sqrt{1+\frac{b^2}{2 a^2}\frac{l^2}{(|l|+2n)^2}} }\,.
\end{equation}
    In our computation the fermionic correlator $\mathcal{C}^\mathrm{fer}(\tau,\sigma)$ is determined only up to terms that are annihilated by the derivatives in~\eqref{eq:bdtdsbf}. We have chosen these ambiguous terms such that $\mathcal{C}^\mathrm{fer}(\tau,\sigma)$ agrees\footnote{Note that in~\eqref{eq:CbosCfer} we have not included the disconnected contribution to the correlator; this contribution can be computed in the free theory and is given by the $O(N)$ term in~\eqref{eq:fercorr} at all values of $b^2/a_0^2$.} up to terms of order $O(b^2)$ with the correlator computed in~\cite{Galliani:2017jlg}. In order to verify that the $O(b^2)$ expansion of the $\mathcal{C}^\mathrm{bos}(z,{\bar z})$ and $\mathcal{C}^\mathrm{fer}(z,{\bar z})$ above agrees with the result obtained via the perturbative method in~\eqref{eq:boscorr1} and~\eqref{eq:fercorr} one can start by expanding each term of the series for small $b$ at fix $a_0$ up to order $b^2$
\begin{equation}
\label{eq:b1sumlnb2}
\mathcal{C}^\mathrm{fer}(\tau,\sigma) \sim
\sum_{l\in\mathbb{Z}} e^{il\sigma}\sum_{n=1}^\infty e^{i l \sigma}e^{i (|l|+2n) \tau} \left[1 + \frac{b^2}{2 a_0^2} \left(-\frac{1}{2} - \frac{l^2}{2(|l|+2 n)^2} + \frac{2 i \tau (|l|+n) n}{|l|+ 2 n}\right) \right]\;.
\end{equation}
The terms in the round parenthesis can be written as ratios of polynomials in the combinations $l$ and $|l|+2n$ that appear in the exponentials. Then it is possible to reduce the sums over $l$ and $n$ in terms of derivative or integrals (with respect to $\tau$ and $\sigma$) of the geometric series. In particular, the presence in the denominator of a factor of $(|l|+2 n)^2$ implies that we have to integrate twice with respect to $\tau$. It is easy to see that the first integration yields logarithms and the second one dilogarithms, producing exactly the terms proportional to ${\rm Li}_2$ in the ${\hat D}$ function present in~\eqref{eq:fercorr}. With some patience it is possible to check that also all other terms of~\eqref{eq:fercorr} are reproduced by performing the sums for the remaining terms in~\eqref{eq:b1sumlnb2}.

\section{CFT interpretation of the bulk correlator}
\label{sec:OPE}

A natural way to make contact with the CFT interpretation is to study the OPE limits. For instance the leading terms of the $z, \,\bar{z}\to 1$ limit (corresponding to the OPE where the two light operators are close) do not receive contributions\footnote{It is easy to see this from~\eqref{eq:Dfrec} by rewriting $\partial_{|z_{12}|^2}$ in terms of $\partial_z$ and $\partial_{\bar{z}}$ and checking that each Jacobian brings a factor of $|1-z|^2$.} from the $\hat{D}_{pp22}$ with $p>1$. By using the definition of appendix~\ref{sec:AppC}, it is straightforward to check that, in this OPE limit, the singular terms obtained from the round parenthesis in~\eqref{eq:boscorr1} and~\eqref{eq:fercorr} are
\begin{equation}
  \label{eq:zto1a}
  \left(-\frac{1}{2}\frac{1}{|1-z|^2}+\sum_{p=1}^k \frac{|z|^2 \hat{D}_{pp22}}{\pi\,p}\right) \sim - \frac{1}{4 (1-z)} - \frac{1}{4 (1-\bar{z})}
\end{equation}
and so do not contribute to the bosonic correlator~\eqref{eq:boscorr1}. The two singular terms above capture the contributions to the fermionic correlator of the $SU(2)_R$ and $SU(2)_L$ currents. After substituting the result~\eqref{eq:zto1a} in~\eqref{eq:fercorr}, we can easily extract the contribution due to the exchange of the $SU(2)_L$ current by focusing on the term proportional to $1/(1-\bar{z})$
\begin{equation}
  \label{eq:tJconf}
  \mathcal{C}^\mathrm{fer}_{{\cal O}(b^2)} \sim \frac{1}{1-\bar{z}} \left[\frac{1}{2} - \frac{1}{4} \sum_k \frac{b_k^2}{a_0^2}\right] = \frac{a^2}{2 a_0^2}\, \frac{1}{1-\bar{z}}~, 
\end{equation}
where in the last line we used~\eqref{eq:constraint}. This provides a check of the relative normalization between the free contribution and the terms proportional to $b_k^2$: at order $1/(1-\bar{z})$ the two combine to produce a result proportional to $a^2$ which is related to the number of strands with ${j}=1/2$. This is the only type of strands in the state considered in section~\ref{sec:bulkc} that can contribute to the exchange of the $SU(2)_L$ currents; in particular, the OPE~\eqref{eq:tJconf} is saturated by the exchange of $J^3$ and, since the correlator factorizes into two protected 3-point functions $\langle  O_H \bar{O}_H J^3 \rangle ~ \langle J^3  O_L \bar{O}_L\rangle$, it is straightforward to check also the overall normalization just by using the free theory result for the 3-point building blocks.

It is possible to extend the result above and focus on the leading term in the $(1-\bar{z})$ expansion, but keep all corrections in $(1-z)$. In Minkowskian signature this corresponds to a light-cone OPE where $y\to t$. Also in this case, only the terms proportional to $\hat{D}_{1122}$ are relevant and we obtain
\begin{equation}
  \label{eq:zbto1}
  {\cal C}^{\rm bos}_{{\cal O}(b^2)} \sim \frac{1}{|1-z|^4} \left\{1- \sum_k \frac{b_k^2}{a_0^2}\left[1+ \frac{1}{2} \frac{1+z}{1-z} \ln z\right]\right\}\;.
\end{equation}
It is interesting to compare this result with the contribution of the (holomorphic) Virasoro block of the identity, but this has to be done with some care. While the heavy operators have conformal weight $h_H=\bar{h}_H=c/24$ (being RR ground state), it is convenient to factor out the contribution of the Sugawara part of the stress tensor that is due to the $SU(2)_L\times SU(2)_R$ R-currents. The reason for doing this is the following: it is possible to take linear combinations of a Virasoro descendant (such as $L_{-2} |0\rangle$) and an affine descendant constructed with the Sugawara stress-tensor (such as $L_{-2}^{\rm Sug} |0\rangle$) to construct a Virasoro primary (i.e. a state annihilated by $L_n$ for $n>0$). So, if we try to interpret the correlators~\eqref{eq:boscorr1} and~\eqref{eq:fercorr} in terms of the full Virasoro blocks, primaries such as the ones mentioned above would appear as new ``dynamical'' contributions. However, their contributions is completely fixed by the symmetries of the theory and so it is more convenient to analyze the bulk results above in terms of the Virasoro blocks generated by $L^{[0]} = L - L^{\rm Sug}$ times the blocks generated by the R-symmetry currents. This approach is particularly apt for the bosonic correlator~\eqref{eq:boscorr1}, since it is not constrained by the R-symmetry at all. By indicating with a superscript $[0]$ all quantities after factoring out the Sugawara contributions, we have 
$h_L^{[0]}=\bar{h}_L^{[0]}=1$ and\footnote{To be precise, the heavy operators dual to the 2-charge geometries are linear combinations of terms with different values of $ h_H^{[0]}$ and $\bar{h}_H^{[0]}$~\cite{Skenderis:2006ah,Kanitscheider:2007wq}. It is possible to calculate the contribution of each term to the correlator as done for instance in~\cite{Giusto:2015dfa} for the 3-point functions, but the result at order $b^2$ coincides with that of the term with the average number of $j=\bar j=1/2$ strands.}
\begin{equation}
  \label{eq:h0}
  h_H^{[0]}=\bar{h}_H^{[0]}=\frac{N}{4} - \frac{\langle J^2 \rangle}{N} = \frac{N}{4} \left[1 - \left(\frac{N^{\left(++\right)}_{1}}{N}\right)^{\!\!\!\! 2}\,\right] \;, 
\end{equation}
where $J^2$ is the Casimir operator of the $SU(2)_L$ algebra and in our case, is sensitive just to the strands with $j,\bar{j}\not=0$. Thus we should compare~\eqref{eq:zbto1} with the contribution of the HHLL identity Virasoro block with the $h_H^{[0]}$ and $h_L^{[0]}$ above, and $c \sim 6 N$ (since subtracting the Sugawara sector does not change the leading $N$ contribution of the D1D5 CFT). By using the results of~\cite{Fitzpatrick:2015zha}, we have that  the leading term in $(1-\bar{z})$ expansion of the leading $N$ contribution of such Virasoro block reads
\begin{equation}
  \label{eq:Idc}
    {\cal C}^{\rm bos}_{{\rm Id}} \sim  \frac{1}{(1-\bar{z})^2} \left[z^{\alpha-1} \left(\frac{\alpha}{1-z^\alpha}\right)^2 \right] \sim \frac{1}{|1-{z}|^4}  \left\{1- \sum_k \frac{b_k^2}{a_0^2}\left[1+ \frac{1}{2} \frac{1+z}{1-z} \ln z\right]\right\}\;,
\end{equation}
where in the second step we used
\begin{equation}
  \label{eq:alphad}
  \alpha = \sqrt{1 - \frac{24  h_H^{[0]}}{c}} = \frac{N^{\left(++\right)}_{1}}{N} = \frac{a^2}{a_0^2} = 1- \sum_k \frac{b_k^2}{2 a_0^2}
\end{equation}
and took the approximation $b_k^2\ll a_0^2$ up to the order $b_k^2/a_0^2$. This shows that the light-cone OPE~\eqref{eq:zbto1} of the strong coupling correlator~\eqref{eq:boscorr1} is entirely saturated by the $L^{[0]}$ Virasoro descendants of the identity~\eqref{eq:Idc}, at least in the ${\cal O}(b^2)$ approximation. Of course the full correlator away from the light-cone limit receives contributions from other $L^{[0]}$ Virasoro blocks. By expanding~\eqref{eq:boscorr1} for $z\to 1$ and $\bar z\to 1$ and comparing with the same expansion of the (left times right) identity Virasoro block, one sees that the first primaries beyond the identity that appear in the OPE have conformal dimension $h={\bar h}=2$. As we argued in the introduction these primaries should be multi-particle operators.

In the case of the heavy state discussed in section~\ref{sec:exb1}, it is possible to show that light-cone OPE reproduces the $L^{[0]}$ identity Virasoro block even at finite values of $b$. Consider first the fermionic correlator in~\eqref{eq:b1sumln}.
The light-cone OPE is captured by the modes with $l\gg n$, so we can approximate each term in the series~\eqref{eq:b1sumln} by expanding the square roots and by neglecting all terms proportional to $1/l$; then, when $z^\alpha$ is not too close to $1$, the leading contribution in the $\bar{z}\to 1$ limit is captured by
\begin{equation}
  \label{eq:sqexp}
\mathcal{C}^\mathrm{fer}(\tau,\sigma)\sim \frac{a^2}{a_0^2} \sum_{l=0}^\infty e^{i l\,(\sigma- \tau)}\,\sum_{n=1}^\infty   e^{-2\,i\, \frac{a^2}{a_0^2} n \tau}=\alpha \,\frac{1}{1-\bar z}\,\frac{1}{1-|z|^{2\alpha}}\,.
\end{equation}
By inserting this approximation in~\eqref{eq:CbosCfer} we have
\begin{equation}
  \label{eq:bexOPE}
  \mathcal{C}^\mathrm{bos}(z,{\bar z}) \sim \partial {\bar \partial} \,\left ( \frac{1}{1-\bar{z}} \frac{\alpha}{1-|z|^{2 \alpha}} \right) \sim  \frac{1}{(1-\bar{z})^2}\, z^{\alpha-1} \left(\frac{\alpha}{1-z^\alpha}\right)^2\,,
\end{equation}
where we focused just on the leading contribution in the limit $\bar{z}\to 1$. As mentioned above, this result agrees with~\eqref{eq:Idc} even at finite values of $b_1$.

\section{Late time behaviour of the exact correlator}
\label{sec:ltb}

For finite $b$ we were not able to resum the series in~\eqref{eq:b1sumln}. However it is still possible to extract useful informations already from~\eqref{eq:b1sumln}, and in particular one can analyze the behaviour of the correlator for large values of the Lorentzian time $\tau$. The aim is to compare the late-time behaviour of the correlator in a pure heavy state with that of the correlator in the naive D1D5 geometry
\begin{equation}\label{eq:BTZ}
ds^2 = \sqrt{Q_1 Q_5}\left[\frac{d r^2}{r^2}+\frac{r^2}{a_0^2}(-d\tau^2 + d\sigma^2)\right]\,,
\end{equation}
which is the limit of the BTZ black hole when both the left and right temperatures are vanishing, and represents the dual of the statistical ensemble of the  RR ground states. Following~\cite{Balasubramanian:2005qu}, it is convenient to focus on the correlator of the two bosonic operators~\eqref{eq:Obos} in this geometry divided by the vacuum 2-point function and the result is
\begin{equation}\label{eq:GBTZ}
\mathcal{G}^\mathrm{bos}_\mathrm{BTZ}(\tau,\sigma)=\frac{1}{4 (\sigma_+-\sigma_-)^2}\left[{\sin^{2}\frac{\sigma_+}{2}}+{\sin^{2}\frac{\sigma_-}{2}}-\frac{4\sin\frac{\sigma_+-\sigma_-}{2} \sin\frac{\sigma_+}{2} \sin\frac{\sigma_-}{2}}{(\sigma_+-\sigma_-)}\right]\,,
\end{equation}
where
$\sigma_\pm \equiv \sigma\pm \tau\,$.
For large $\tau$ this correlator vanishes like
\begin{equation}\label{eq:decay}
\mathcal{G}^\mathrm{bos}_\mathrm{BTZ}(\tau,\sigma)\sim \frac{1}{\tau^2}\,.
\end{equation}
This large-time decay is a signal of information loss~\cite{Maldacena:2001kr}: the decay in~\eqref{eq:decay} is polynomial rather than exponential, because the naive geometry~\eqref{eq:BTZ} is a degenerate zero-temperature limit of a regular finite-temperature black hole. 

Let us now consider the correlator in the pure heavy state characterized by $b_k=b \delta_{k,1}$ studied in section~\ref{sec:exb1}. The result of the previous section implies that, for generic values of $\sigma=\sigma_0$, the correlator given in~\eqref{eq:b1sumln} has the same singularities at $\tau_k=\sigma_0 + 2\pi k$ as the vacuum correlator. Indeed in this regime the leading contribution to the sum comes from the modes with $l\gg n$ and so, close to $\tau_k$ the fermionic and bosonic correlators are well approximated by~\eqref{eq:sqexp} and~\eqref{eq:bexOPE}. Then, as expected for a pure state, we have that $\mathcal{G}^\mathrm{bos}_{b_1}$ or $\mathcal{G}_{b_1}^\mathrm{fer}$ tend to a finite value when $\tau\to\tau_k$ for every $k$:
\begin{equation}\label{eq:oscillations}
\mathcal{G}^\mathrm{fer}_{b_1}\sim \alpha \frac{1-e^{2 i \sigma_0}}{1-e^{2i \alpha \sigma_0} e^{2\pi i \alpha k}}\,,\quad \mathcal{G}^\mathrm{bos}_{b_1}\sim \alpha^2 e^{2i \sigma_0 (\alpha-1)} e^{2\pi i \alpha k} \left(\frac{1-e^{2 i \sigma_0}}{1-e^{2i \alpha \sigma_0} e^{2\pi i \alpha k}}\right)^2\,.
\end{equation}
This is in contrast with what happens in the case of the naive geometry~\eqref{eq:decay} where $\mathcal{G}^\mathrm{bos}_\mathrm{BTZ}$ goes to zero at late times.

Since the geometries \eqref{eq:solb}, dual to the pure states \eqref{eq:ensembleD1D5}, reduce to the naive D1D5 geometry \eqref{eq:BTZ} in the limit $a\to 0$, it is interesting to ask if the non-unitary correlator \eqref{eq:GBTZ} emerges as the $a\to 0$ limit of the pure state correlator (\ref{eq:CbosCfer},\ref{eq:b1sumln}). When $a\ll b$, one can distinguish two contributions to the series in \eqref{eq:b1sumln}: 
\begin{subequations}
\begin{align}
\frac{a_0}{a} |l| \gg 2 n &:\quad \mathcal{C}^\mathrm{fer}\sim \frac{a^2}{a_0^2}\sum_{l,n}\left(1+\frac{2 n}{|l|}\right)e^{i (l \sigma -  |l| \tau)}\, ;\\
\frac{a_0}{a} |l| \ll 2 n &:\quad \mathcal{C}^\mathrm{fer}\sim \frac{a}{a_0}\sum_{l,n}e^{i l \sigma} e^{-i \frac{a}{a_0} 2 n \tau}\,,
\end{align}
\end{subequations}
where we have used that $\frac{a}{a_0}\sim \frac{\sqrt{2} a}{b}$. The terms in the first line of the equation above give the sum of a function of $\sigma+\tau$ and a function of $\sigma-\tau$, and hence do not contribute to the bosonic correlator. We thus keep only the second type of contributions, which give
\begin{equation}\label{eq:corrasmall}
\begin{aligned}
 \mathcal{C}^\mathrm{fer}(\sigma,\tau)&\sim  \frac{a}{a_0}\sum_{l\in \mathbb{Z}}e^{i l \sigma} \sum_{n=\frac{a_0}{2 a}|l|}^\infty e^{-i \frac{a}{a_0} 2 n \tau}+\ldots\\
 &=\frac{a}{a_0}\frac{1}{1-e^{- 2 i \frac{a}{a_0} \tau}}\left[ \frac{1}{1-e^{i(\sigma-\tau)}}+ \frac{1}{1-e^{-i(\sigma+\tau)}}-1\right]+\ldots\,,
\end{aligned}
\end{equation}
where the dots are the terms that do not contribute to $\mathcal{C}^\mathrm{bos}$. No matter how small $a/a_0$ is, as far as $a$ is non-zero the correlator in \eqref{eq:corrasmall} and the bosonic correlator derived from it have an oscillating non-vanishing behaviour for large enough $\tau$, as was found\footnote{Note however that one cannot directly compare the $\tau\to \sigma$ limit of \eqref{eq:corrasmall} with \eqref{eq:oscillations} or with \eqref{eq:sqexp}, because both results are not valid when $z^\alpha$ is close to 1.} in \eqref{eq:oscillations} for finite $a$. However, if one observes the correlators at times $\tau\ll a_0/a$, one can approximate \eqref{eq:corrasmall} as
\begin{equation}
\mathcal{C}^\mathrm{fer}(\sigma,\tau)\sim \frac{1}{2 i \tau}\left[ \frac{1}{1-e^{i(\sigma-\tau)}}+ \frac{1}{1-e^{-i(\sigma+\tau)}}-1\right]+\ldots\,,
\end{equation}
and one obtains precisely the ``naive'' correlator given in \eqref{eq:GBTZ}. We conclude that the correlator in the naive geometry \eqref{eq:BTZ} approximates the correlators in pure states in the limit $a\ll a_0$ and for times $\tau$ shorter than $a_0/a$.

\section{Summary and outlook}
\label{sec:outlook}

In this paper we used the supergravity approximation of type IIB string theory to derive, via the AdS$_3$/CFT$_2$, the strong coupling expression for the HHLL correlators~\eqref{eq:CcalG} where the two light operators are the bosonic states in~\eqref{eq:Obos} and the heavy operators belong to the ensemble of RR ground states in~\eqref{eq:ensembleD1D5}. As reviewed in Section~\ref{sec:orbCFT}, at the orbifold point in the superconformal moduli space, it is straightforward to calculate these correlators in full generality. This was exploited in~\cite{Balasubramanian:2005qu,Balasubramanian:2016ids} to extract interesting properties of the correlators for {\em generic} RR ground states. Of course in order to study the problem in a regime where weakly coupled AdS gravity is a valid approximation, one needs to deform the orbifold description and move to a region where the CFT is strongly coupled. Here we bypassed this challenging task by working directly with the supergravity description, and to make the computation feasible we restricted to the regime ($N_k^{(0)}\ll N_{1}^{\left(++\right)}$) where the states are close to the RR ground state with maximal R-charge. For a particular family of states (with $N_k^{(0)}=0$ for $k\ge 2$) we were able to compute the correlator at strong coupling for all values of the R-charge (even if only in the form of a Fourier series), including the limit in which the R-charge becomes vanishingly small. To make contact between the gravity results (\eqref{eq:boscorr1}, \eqref{eq:fercorr} and \eqref{eq:CbosCfer}, \eqref{eq:b1sumln}) and the CFT point of view, we started to look at different OPE limits of the correlator. In the light-cone OPE limit the only contributions to the bosonic correlator come from the Virasoro descendants of the identity, as expected (see for instance the discussion in Appendix A of \cite{Fitzpatrick:2016mjq}) for generic correlators in a CFT where the stress tensor is the only conserved current. In the usual Euclidean OPE, however, other primaries beyond the identity contribute, the first ones appearing at dimension $h=\bar h=2$ for the bosonic correlator. Summing over these primaries crucially changes the qualitative late time behaviour of the correlator: while each individual classical Virasoro conformal block vanishes at late times, we verify in Section~\ref{sec:ltb} that our correlator has an oscillatory behaviour for arbitrarily large time, as expected in a unitary theory without information loss. Note that this results holds also for states that are far from the maximally spinning ground state, for which the correlator is dynamical and not fixed by the symmetries.

We thus see that correlators in pure states are consistent with unitarity both at the orbifold and at the supergravity point, but the exchanged operators that guarantee the unitary behaviour are different at the two points. While in the free theory correlators receive contributions from an infinite series of conserved currents that are lifted at a generic point in the moduli space, contributions from new primaries appear in the strong coupling result. Since, in all known cases, non-protected single trace operators acquire divergent anomalous dimensions in the supergravity limit, these primaries must be multi-particle operators, i.e. operators made by products of fields evaluated on different copies of the CFT. Multi-particle operators generically have anomalous dimensions and three-point functions that acquire moduli-dependent corrections in the $1/N$ expansion, and hence they can give finite contributions to correlators that are not visible at the orbifold point. One of the most immediate and potentially interesting developments of our work is a closer analysis of these multi-particle operators. An extension of the techniques developed in the AdS$_5$/CFT$_4$ context (see for instance \cite{D'Hoker:1999jp,Aprile:2017xsp}) should allow us to extract the first corrections to the anomalous dimensions and the three-point functions from the supergravity correlators, thus investigating the consistency of our results and gaining a better understanding of the mechanism by which information is encoded in the dynamical correlators. 

The analysis in this article has been limited to RR ground states, for which we have complete control over the dual supergravity geometries. Though these states have interesting statistical properties and an entropy that scales like a positive power of the central charge, they represent a ``degenerate'' toy model for a black hole, in the sense that the ensemble of such states is not described by a black hole with a finite horizon in classical supergravity. It would thus be significant to extend our analysis to states with an excited left (or the right) sector. In particular a family of such states has been recently contsructed \cite{Bena:2015bea,Bena:2016agb,Bena:2016ypk}, of which a subset is known\cite{Bena:2017upb} to have factorizable 6D metric, in the sense explained in Section~\ref{sec:bulkc}. It would be interesting to see if the general mechanism for information conservation suggested by our study is confirmed in an ensemble dual to a regular black hole, or if new qualitative features emerge.

\vspace{7mm}
 \noindent {\large \textbf{Acknowledgements} }

 \vspace{5mm} 

We would like to thank I. Bena, M. Guica, E. Martinec, M. Shigemori, D. Turton and N. Warner for discussions and correspondence. 
S.G. and R.R. wish thank the Galileo Galilei Institute for Theoretical Physics (GGI) for the hospitality during the program ``New Developments in AdS3/CFT2 Holography''. This research is partially supported by STFC (Grant ST/L000415/1, {\it String theory, gauge theory \& duality}). 

\appendix

\section{The orbifold D1D5 CFT}
\label{sec:AppB}

For the orbifold D1D5 CFT we follow the conventions of~\cite{Galliani:2016cai}. In particular, in deriving the Ward identity~\eqref{eq:WIg} we used the explicit form of the left and right supercurrents
\begin{equation}
\label{eq:GtG}
G^{\alpha}_{A}(z) \equiv \sum_{r=1}^N \partial X_{A\dot{A}\,(r)} \psi^{\alpha \dot{A}}_{(r)}\;,\qquad \tilde{G}^{\dot{\alpha}}_{A}(z) \equiv \sum_{r=1}^N \bar{\partial} X_{A\dot{A} (r)} \tilde{\psi}^{\dot{\alpha} \dot{A}}_{(r)}\;,
\end{equation}
and the OPE between the elementary fields
\begin{equation}
  \label{eq:elopes}
  \psi^{\alpha\dot{A}}_{(r)}(z)\,\psi^{\beta\dot{B}}_{(s)}(w) \sim - \frac{\epsilon^{\alpha\beta}\,\epsilon^{\dot{A}\dot{B}}\,\delta_{r,s}}{z-w}  
~,~~~
\partial X^{A\dot{A}}_{(r)}(z)\,\partial X^{B\dot{B}}_{(s)}(w) \sim \frac{\epsilon^{AB}\epsilon^{\dot{A}\dot{B}}\,\delta_{r,s}}{(z-w)^2} 
\end{equation}
where the $SU(2)$ indices are raised and lowered by using the $\epsilon$ tensor with the convention $\epsilon_{12} = - \epsilon_{21} = \epsilon^{21} = -\epsilon^{12} = +1$, for instance
\begin{equation}
\partial X_{A\dot{A}} = \epsilon_{AB} \epsilon_{\dot{A}\dot{B}}\,\partial X^{B\dot{B}},\qquad \partial X^{A\dot{A}} = \epsilon^{AB} \epsilon^{\dot{A}\dot{B}}\,\partial X_{B\dot{B}}
\end{equation}
and similarly for the antiholomorphic fields.

In a twisted sector, the boundary conditions mix different copies of the CFT to form a strand of length $k$, which means that we have the following periodicites
\begin{equation}
\label{eq:bck}
\partial X^{A\dot{A}}_{(r)}\left( e^{2\pi\ii} z \right) = \partial X^{A\dot{A}}_{(r+1)}(z),\qquad \bar{\partial} X^{A\dot{A}}_{(r)}\left( e^{-2\pi\ii} \bar{z} \right) = \bar{\partial} X^{A\dot{A}}_{(r+1)}(\bar{z}),
\end{equation}
with the identification $\partial X^{A\dot{A}}_{(k+1)}\equiv \partial X^{A\dot{A}}_{(1)}$ and $\bar{\partial} X^{A\dot{A}}_{(k+1)}\equiv \bar{\partial} X^{A\dot{A}}_{(1)}$ and $r = 1,\ldots, k$. It is possible to diagonalize the boundary conditions by taking linear combinations of the fields for different values of $(r)$. We label the independent fields of this new basis with the index $\rho = 0, \ldots, k-1$,
\begin{subequations}
 \label{eq:bosonsrtorhobasischange}
\begin{align}
\partial X^{1\dot{A}}_\rho (z) &=\frac{1}{\sqrt{k}} \sum_{r=1}^k e^{2\pi\ii \frac{r\rho}{k}}\,\partial X^{1\dot{A}}_{(r)}(z),& \partial X^{2\dot{A}}_\rho (z) &= \frac{1}{\sqrt{k}} \sum_{r=1}^k e^{-2\pi\ii \frac{r\rho}{k}}\,\partial X^{2\dot{A}}_{(r)}(z),\\
\bar{\partial} X^{1\dot{A}}_\rho(\bar{z}) &= \frac{1}{\sqrt{k}} \sum_{r=1}^{k} e^{-2\pi\ii \frac{r\rho}{k}}\,\bar{\partial} X^{1\dot{A}}_{(r)}(\bar{z}),& \bar{\partial} X^{2\dot{A}}_\rho(\bar{z}) &= \frac{1}{\sqrt{k}} \sum_{r=1}^{k} e^{2\pi\ii \frac{r\rho}{k}}\,\bar{\partial} X^{2\dot{A}}_{(r)}(\bar{z}),
\end{align}
\end{subequations}
with the (diagonalized) monodromy conditions in the $\rho$ basis now being
\begin{subequations}
 \label{eq:twisteddiagboundcondbos}
\begin{align}
\partial X^{1\dot{A}}_\rho \left( e^{2\pi\ii} z\right) &= e^{-2\pi\ii \frac{\rho}{k}}\,\partial X^{1\dot{A}}_\rho(z), & \partial X^{2\dot{A}}_\rho \left( e^{2\pi\ii} z\right) &= e^{2\pi\ii \frac{\rho}{k}}\,\partial X^{2\dot{A}}_\rho(z) ,\\
\bar{\partial} X^{1\dot{A}}_\rho(e^{-2\pi\ii}\,\bar{z}) &= e^{2\pi\ii \frac{\rho}{k}}\,\bar{\partial} X^{1\dot{A}}_\rho(\bar{z}),& \bar{\partial} X^{2\dot{A}}_\rho(e^{-2\pi\ii}\,\bar{z}) &= e^{-2\pi\ii \frac{\rho}{k}}\,\bar{\partial} X^{2\dot{A}}_\rho(\bar{z})\;.
\end{align}
\end{subequations}
Then the standard mode expansion following from \eqref{eq:twisteddiagboundcondbos} are
\begin{subequations}
 \label{eq:twistedbosonsmodeexpansion}
\begin{align}
\partial X^{1\dot{A}}_\rho(z) &= \sum_{n\in\mathbf{Z}} \alpha^{1\dot{A}}_{\rho, n+\frac{\rho}{k}} z^{-n-1-\frac{\rho}{k}}, & \partial X^{2\dot{A}}_\rho(z) &= \sum_{n\in\mathbf{Z}} \alpha^{2\dot{A}}_{\rho,n-\frac{\rho}{k}} z^{-n-1+\frac{\rho}{k}},\\
\bar{\partial} X^{1\dot{A}}_\rho(\bar{z}) &= \sum_{n\in\mathbf{Z}}  \tilde{\alpha}^{1\dot{A}}_{\rho,n+\frac{\rho}{k}} \bar{z}^{-n-1-\frac{\rho}{k}}, & \bar{\partial} X^{2\dot{A}}_\rho(\bar{z}) &= \sum_{n\in\mathbf{Z}}  \tilde{\alpha}^{2\dot{A}}_{\rho,n-\frac{\rho}{k}} \bar{z}^{-n-1+\frac{\rho}{k}}\;.
\end{align}
\end{subequations}
Notice that we can use~\eqref{eq:bosonsrtorhobasischange} and rewrite the $k$ terms belonging to a single strand in the operators~\eqref{eq:Obos} as a sum over $\rho$
\begin{equation}
  \label{eq:rtorho}
   \sum_{r=1}^{k} \partial X^{A\dot{B}}_{(r)}(z) \bar{\partial} X^{A\dot{C}}_{(r)}(\bar{z}) =  \sum_{\rho = 0}^{k-1} \partial X^{A\dot{B}}_{\rho}(z) \bar{\partial} X^{A\dot{C}}_{\rho}(\bar{z})\;.
\end{equation}
Then by the commutation relations in the twisted sector
\begin{equation}
 \label{eq:commrel_bos_rho}
\left[ \alpha^{A\dot{A}}_{\rho_1,n}, \alpha^{B\dot{B}}_{\rho_2,m}\right] = \epsilon^{AB} \epsilon^{\dot{A}\dot{B}}\,n\,\delta_{n+m,0}\,\delta_{\rho_1, \rho_2}\;,
\end{equation}
we can easily calculate the 2-point correlator on strand of length $k$
\begin{equation}
\,_k \langle  0| \partial X^{1\dot{1}}_{\rho}(z_1) \, \partial X^{2\dot{2}}_{\rho}(z_2) |0\rangle_k = \frac{1}{(z_1- z_2)^{2}} \left( \frac{z_1}{z_2}\right)^{-\frac{\rho}{k}}  \left\lbrace 1 - \frac{\rho}{k} \left(1-\frac{z_1}{z_2}\right) \right\rbrace,
\end{equation}
with similar formulae holding for the antiholomorphic sector. Then the contribution from such strand to the correlator~\eqref{eq:4corr} with the light operators in~\eqref{eq:Obos} is
  \begin{align}
\label{eq:Gbosz}
\mathcal{C}_k^{\rm bos} (z, \bar{z}) = & \frac{1}{(1-z)^2(1-\bar{z})^2}  \sum_{\rho = 0}^{k-1} \left|{z}\right|^{\frac{2\rho}{k}} \left|1 - \frac{\rho}{k} \left(1- \frac{1}{z}\right)\right|^2
\\\nonumber = & 
\partial \bar\partial\left[\frac{1- z \bar{z}}{(1-z)(1-\bar{z})\left(1-(z \bar{z})^\frac{1}{k} \right)} \right]\;.
  \end{align}
As explained in section~\ref{sec:orbCFT}, the possibility of writing the result as in the second line follows from a Ward identity with a correlator where the light operators are (anti)-chiral primaries. It is also interesting to write the result in terms of $z = e^{-\ii w}$ and $\bar{z} = e^{\ii \bar{w}}$. By including a factor of $e^{-i(w-\bar{w})}$ which follows from the Jacobian necessary to transform the correlator from the plane to the cylinder coordinates, one has
\begin{equation}
\label{eq:Gbosw}
\mathcal{C}_k^{\rm bos} (w, \bar{w}) = \frac{1}{16 k \sin^2 \left(\frac{w-\bar{w}}{2k}\right)} \left[ \frac{1}{\sin^2\left(\frac{w}{2}\right)} + \frac{1}{\sin^2\left(\frac{\bar{w}}{2}\right)} - \frac{2\sin\left( \frac{w-\bar{w}}{2}\right)}{k\tan\left(\frac{w-\bar{w}}{2k}\right) \sin\left(\frac{w}{2}\right)\sin\left(\frac{\bar{w}}{2}\right)}\right]\;.
\end{equation}

By following a similar approach it is straightforward to calculate the contribution of a strand of length $k$ to the correlator with the fermionic light operators~\eqref{eq:Ofer}
\begin{equation}
  \label{eq:Cjbj}
\mathcal{C}_{k\,(j\,\bar{j})}^{\rm fer} = \frac{1}{|z|} \frac{|z|^{\frac{2}{k}}- |z|^2}{(1-z)(1-\bar{z})\left(1-|z|^\frac{2}{k} \right)} + f_{(j,\bar{j})}(z,\bar{z}) \;,
\end{equation}
where $ f_{k\,(j,\bar{j})}$ is the $\rho=0$ contribution which depends on the $SU(2)_L\times SU(2)_R$ quantum numbers
\begin{equation}
\label{eq:fkj}
\begin{aligned}
   f_{(j,\bar{j})} = & \frac{z^j \bar{z}^{\bar j}}{(1-z)(1-\bar{z})}\;,~~~\mbox{with}~j\,,\bar{j}=\pm \frac{1}{2}\;, \\ 
 f_{(0,0)} = & \frac{1}{2 |z| (1-z)(1-\bar{z})} \left(1+ |z|^2+ |1-z|^2 \right)\;.
\end{aligned}
\end{equation}


\section{Wave equation}
\label{sec:appA}

The CFT operator $\partial X^{(i} \bar \partial X^{j)}$, with $i,j=1,\ldots,4$, is dual to a deformation $h_{ij}$ of the $T^4$ metric. For simplicity we restrict here to a traceless deformation $\delta^{ij} h_{ij}=0$. We derive here the linearized equation satisfied by $h_{ij}$ in the background of a generic two-charge microstate. When the background is that of the naive D1D5 geometry, it is know that $h_{ij}$ is a minimally coupled scalar (see for example \cite{David:2002wn}). We show that this remains true for a generic D1D5 microstate.

The deformed 10D string metric is
\begin{equation}
ds^2_{10}= \sqrt{\frac{Z_1 Z_2}{\mathcal{P}}}\,ds^2_6 + \sqrt{\frac{Z_1}{Z_2}}\,(\delta_{ij}+h_{ij})\,dz^i dz^j\,,
\end{equation}
where $\mathcal{P}$ is defined in \eqref{eq:calP} and $ds^2_6$ is the 6D Einstein metric given in \eqref{eq:6DE}. The background solution also contains the dilaton $\Phi$, the RR 1-form $F_1$, the NSNS and RR three-forms $H_3$ and $F_3$ and the self-dual RR 5-form $F_5$:
\begin{subequations}\label{eq:otherfields}
\begin{align}
&e^{2\Phi}= \frac{Z_1^2}{\mathcal{P}}\,,\quad F_1 = d\!\left(\frac{Z_4}{Z_1}\right)\,,\\
&H_3=-d\hat u\wedge d\hat v \,d\!\left(\frac{Z_4}{\mathcal{P}}\right)-\frac{Z_4}{\mathcal{P}}(d\hat v\wedge d\omega-d\hat u\wedge d\beta)+*_4 dZ_4\,,\\
&F_3=\frac{d\hat u\wedge d\hat v}{\mathcal{P}}\left(\frac{Z_2}{Z_1} dZ_1 - \frac{Z_4}{Z_1}dZ_4\right)-\frac{1}{Z_1}(d\hat v\wedge d\omega-d\hat u\wedge d\beta)+*_4 dZ_2-\frac{Z_4}{Z_1}*_4 d Z_4\,,\\
&F_5=-\frac{d\hat u\wedge d\hat v}{\mathcal{P}}\wedge *_4 \left(Z_4 \,dZ_2 - Z_2 \,dZ_4\right)+d\!\left(\frac{Z_4}{Z_2}\right)\wedge dz^1\wedge dz^2 \wedge dz^3\wedge dz^4\,,
\end{align}
\end{subequations}
where for brevity we have denoted 
\begin{equation}
d\hat u\equiv du + \omega \,,\quad d\hat v\equiv dv + \beta\,,
\end{equation}
and $*_4$ is the Hodge dual done with $ds^2_4$.

We would like to derive the equations of motion at first order in $h_{ij}$. The only non-trivial equation is Einstein's equation:
\begin{equation}\label{eq:einstein}
\begin{aligned}
&e^{-2\Phi}\left(R_{MN} + 2 \nabla_M\nabla_N \Phi \right)+\frac{1}{4}g_{MN}\left(F_P F^P+\frac{1}{3!}F_{PQR} F^{PQR}\right)-\frac{1}{4} \frac{1}{4!} F_{MPQRS} {F_N}^{PQRS}\\
&-\frac{1}{2}F_M F_N - \frac{1}{4} e^{-2\Phi} H_{MPQ}{H_{N}}^{PQ}-\frac{1}{2} \frac{1}{2!} F_{MPQ} {F_N}^{PQ}=0\,,
\end{aligned}
\end{equation}
where the Ricci tensor $R_{MN}$, the covariant derivatives and the raising of indices are referred to the string metric; we have omitted to write the subscripts indicating the form degree since the explicit presence of the indices leaves no space to confusion. The second line of \eqref{eq:einstein} does not receive corrections in $h_{ij}$; the first line is non-trivial only when both indices $M$, $N$ are along $T^4$. One finds
\begin{equation}\label{eq:e1}
\begin{aligned}
\delta R_{ij} =&\,-\frac{1}{2}\frac{\sqrt{\mathcal{P}}}{Z_2}\Biggl[\square_6 h_{ij}+\frac{\mathcal{P}}{Z_1^2}\partial^{\mu}\!\left(\frac{Z_1^2}{\mathcal{P}}\right)\partial_{\mu} h_{ij}\\
&+\frac{1}{2}\left(\frac{Z_2}{Z_1} \square_6\! \left(\frac{Z_1}{Z_2}\right)+\frac{\mathcal{P}}{Z_1^2} \partial^\mu \!\left(\frac{Z_1 Z_2}{\mathcal{P}}\right) \partial_\mu\! \left(\frac{Z_1}{Z_2}\right)\right) h_{ij}\Biggr]\,,
\end{aligned}
\end{equation}
\begin{equation}\label{eq:e2}
\delta (\nabla_i \nabla_j \Phi)=\frac{1}{4}\frac{\mathcal{P}^{3/2}}{Z_1^2 Z_2} \partial^{\mu}\!\left(\frac{Z_1^2}{\mathcal{P}}\right)\left[\partial_\mu h_{ij}+\frac{1}{2}\frac{Z_2}{Z_1}\partial_\mu\!\left(\frac{Z_1}{Z_2}\right) h_{ij}\right]\,,
\end{equation}
\begin{equation}\label{eq:e3}
\begin{aligned}
F_P F^P+\frac{1}{3!}F_{PQR} F^{PQR}=&\,\frac{\sqrt{\mathcal{P}}}{Z_1  Z_2^2}\,\Biggl[\partial^\mu Z_2 \partial_\mu Z_2 - \frac{\mathcal{P}Z_2}{Z_1^3}\partial^\mu Z_1 \partial_\mu Z_1 \\
&+\frac{Z_2}{Z_1}\partial^\mu Z_4 \partial_\mu Z_4 -2 \frac{Z_4}{Z_1}\partial^\mu Z_2 \partial_\mu Z_4 \Biggr]\,,
\end{aligned}
\end{equation}
\begin{equation}\label{eq:e4}
 \frac{1}{4!} \delta(F_{iPQRS} {F_j}^{PQRS})=\frac{\sqrt{\mathcal{P}} Z_2}{Z_1^2}\,\partial^\mu\!\left(\frac{Z_4}{Z_2}\right)\partial_\mu\!\left(\frac{Z_4}{Z_2}\right)\,h_{ij}\,,
\end{equation}
and of course $\delta g_{ij} =  \sqrt{\frac{Z_1}{Z_2}}\,h_{ij}$. Here $\square_6$ is the scalar laplacian of the 6D Einstein metric $ds^2_6$ and the 6D indices $\mu$ are raised and lowered with $ds^2_6$. The warp factors $Z_1$ and $Z_2$ of a generic two-charge microstate are harmonic: $\square_6 Z_1 = \square_6 Z_2=0$. Exploiting this property, the variation of the first two terms of \eqref{eq:einstein} can be simplified to
\begin{equation}\label{eq:e1plus2}
\begin{aligned}
e^{-2\Phi} \left[ \delta R_{ij} + 2 \delta (\nabla_i \nabla_j \Phi) \right]  =   - \frac{1}{4}\, \frac{\mathcal{P}^{3/2}}{Z_1^3 Z_2} \left[ 2 Z_1 \square_6 h_{ij} + \!\left(\frac{Z_1}{Z_2^2} \,\partial_\mu Z_2 \partial^\mu Z_2 - \frac{1}{Z_1} \, \partial_\mu Z_1 \partial^\mu Z_1\right) \!h_{ij} \right] .
\end{aligned}
\end{equation}
Substituting \eqref{eq:e1plus2}, \eqref{eq:e3} and \eqref{eq:e4} in the first line of \eqref{eq:einstein} one can verify that at first order in $h_{ij}$ the equation reduces to 
\begin{equation}
\square_6 h_{ij}=0\,,
\end{equation}
i.e. $h_{ij}$ is a minimally coupled scalar in 6D. 


\section{Bulk integrals}
\label{sec:AppC}
We describe here the steps that lead from \eqref{eq:integral} to \eqref{eq:boscorr1}. The manipulations we perform are standard in Witten diagrams computations and are similar to the ones described in Appendix  E of \cite{Galliani:2017jlg}. 

The first term of the source $\langle J_k \rangle$ in \eqref{eq:source} can be conveniently rewritten as
\begin{equation}
-\frac{r}{(r^2+a_0^2)} \partial_r B_0=\frac{1}{2 a_0^2}(B_-\partial_\mu B_+ + B_+ \partial_\mu B_-)\partial^\mu B_0\,,
\end{equation}
where we have introduced
\begin{equation}
B_{\pm}\equiv \frac{a_0}{\sqrt{r^2+a_0^2}}\,e^{\pm  t_e/R}\,,
\end{equation}
and it is understood that indices are raised and lowered with the Euclidean version of the AdS$_3$ metric $g_{AdS_3}$. $B_+$ and $B_-$ are the bulk-to-boundary propagators with $\Delta=1$ evaluated at the points $z=\infty$ and $z=0$. 
 It is convenient to start from the version of \eqref{eq:integral} written on the Euclidean cylinder:
\begin{equation}\label{eq:corrcyl}
\begin{aligned}
&\langle O_H(t_e=\!-\infty) {\bar O}_H(t_e=\!\infty) O_L(0,0) {\bar O}_L(t_e,y) \rangle |_{b_k^2}=\\
&=-\sum_k \frac{b_k^2}{2\pi} \int \! d^3 {\bf r}'_e \sqrt{\bar g} \,K_2^\mathrm{Glob}({\bf r}'_e|t_e,y)\,\langle J_k({\bf r}'_e)\rangle 
= -\sum_k \frac{b_k^2}{2\pi a_0^{2}}\left(\frac{I_1+I_2}{2}- I_3-\sum_{p=2}^k \frac{1}{2\,p}\tilde I_p\right)\,,
\end{aligned}
\end{equation}
where
\begin{subequations}
\begin{align}
I_1&\equiv \int \!d^3 {\bf r}'_e \sqrt{\bar g} \,B_0({\bf r}'_e|t_e,y)\,\partial'^{\mu}B_0({\bf r}'_e|0,0)\,B_-({\bf r}'_e)\,\partial'_\mu B_+({\bf r}'_e)\,,\\
I_2&\equiv \int \!d^3 {\bf r}'_e \sqrt{\bar g} \,B_0({\bf r}'_e|t_e,y)\,\partial'^{\mu}B_0({\bf r}'_e|0,0)\,B_+({\bf r}'_e)\,\partial'_\mu B_-({\bf r}'_e)\,,\\
I_3&\equiv \int \!d^3 {\bf r}'_e \sqrt{\bar g} \,B_0({\bf r}'_e|t_e,y)\,R^2 \partial^2_{t'_e}B_0({\bf r}'_e|0,0)\,\frac{a_0^4}{(r'^2+a_0^2)^2}\,,\\
\tilde I_p&\equiv \int \!d^3 {\bf r}'_e \sqrt{\bar g} \,B_0({\bf r}'_e|t_e,y)\,R^2(\partial^2_{t'_e}+\partial_{y'}^2)B_0({\bf r}'_e|0,0)\,\frac{a_0^{2p}}{(r'^2+a_0^2)^p}\,.
\end{align}
\end{subequations}
These integrals can be written in terms of the same $D$-functions $D_{p_1 p_2 p_3 p_4}$ that appear in the computations of Witten's diagrams. The $D$-functions that we need in this paper can be computed by starting from 
\begin{equation}
  \label{eq:Df}
  D_{1111}(z_1,z_2,z_3,z_4) = \frac{\pi}{|z_{13}|^2 |z_{24}|^2 (z-\bar{z})} \left( {\rm Li}_2(z) - {\rm Li}_2(\bar{z}) + \ln|z| \ln\frac{1-z}{1 - \bar{z}} \right)\;,
\end{equation}
where $z_{kl} = z_k-z_l$ and $z$ is given in~\eqref{eq:zcr}. Each pair $(kl)$ of subscripts can be increased by one by taking the derivative with respect to the corresponding $|z_{kl}|^2$; hence one has
\begin{equation}
  \label{eq:Dfrec}
  {D}_{p_1+1\, p_2+1\, p_3\, p_4} = - \frac{\hat{p}-d}{2 p_1 p_2} \frac{\partial}{\partial |z_{12}|^2} D_{p_1\, p_2\, p_3\, p_4}
\end{equation}
and its permutations (with $\hat{p}=\sum_i p_i$ and, in our case, $d=2$). It is also convenient to introduce the rescaled functions 
\begin{equation}
  \label{eq:Dhf}
  \hat{D}_{p_1 \, p_2\, p_3\, p_4} = \lim_{z_2\to\infty} |z_2|^{2p_2} D_{p_1 \,p_2\, p_3\, p_4}(0,z_2,1,z)\,.
\end{equation}

As explained\footnote{With respect to \cite{Galliani:2017jlg}, we have renamed the integrals and the bulk-to-boundary propagator $B_0$ has now $\Delta=2$, instead on $\Delta=1$. The definition of the functions $\hat D$ is the same as given in eq. (D.2) of \cite{Galliani:2017jlg}.} around (E.10) of \cite{Galliani:2017jlg}, one has
\begin{equation}
I_1 + I_2 = 2 |z|^2\hat D_{2222}\,.
\end{equation}
$I_1$ can be computed as in (E.8) of \cite{Galliani:2017jlg} by writing the integral in Poincar\'e coordinates $\mathbf{w}\equiv \{w_0,w,\bar w\}$:
\begin{equation}
\begin{aligned}
|z|^{-2} I_1 =&\, \int \! d^3\mathbf{w}\,w_0^{-1} \left(\frac{w_0}{w_0^2+|w-z|^2}\right)^2 \partial_{w_0}\left(\frac{w_0}{w_0^2+|w-1|^2}\right)^2\frac{w_0}{w_0^2+|z|^2}\\
=&\, \int \! d^3\mathbf{w}\,w_0^{-1} \left(\frac{w_0}{w_0^2+|w-z|^2}\right)^2 \left[ \frac{2 w_0}{(w_0^2 + |w-1|^2)^2}- \frac{4 w_0^3}{(w_0^2 + |w-1|^2)^3}\right] \frac{w_0}{w_0^2+|z|^2}\\
=&\,2 \hat D_{1122}-4 \hat D_{1232}\,.
\end{aligned}
\end{equation}
Therefore
\begin{equation}
|z|^{-2} I_2 = 2 \hat D_{2222}-2 \hat D_{1122}+4 \hat D_{1232}\,. 
\end{equation}
The computation of $I_3$ follows (E.14):
\begin{equation}\label{eq: identity}
\begin{aligned}
I_3 &= R \partial_{t_e} \frac{I_1-I_2}{2} = (z \partial + \bar z \bar \partial)\left(|z|^2(2 \hat D_{1122}-4 \hat D_{1232}- \hat D_{2222})\right)\\
&= \frac{2 |z|^2}{|1-z|^4}\left(2 (1+|z|^2) \hat D_{3311}-\pi \right)\,,
\end{aligned}
\end{equation}
where the last identity follows from a computation that uses the explicit expression of the $\hat D$-functions. Finally
\begin{equation}
\begin{aligned}
\tilde I_p &=R^2(\partial^2_{t_e}+\partial_{y}^2) \int \!d^3 {\bf r}'_e \sqrt{\bar g} \,B_0({\bf r}'_e|t_e,y) B_0({\bf r}'_e|0,0)\,\frac{a_0^{2p}}{(r'^2+a_0^2)^p}\\
&=4 \partial \bar \partial (|z|^2 \hat D_{pp22})\,.
\end{aligned}
\end{equation}
Substituting the above expressions for the integrals in \eqref{eq:corrcyl}, transforming to the Euclidean plane and adding the trivial contribution $1/|1-z|^4$ from $b_k=0$, one finds the correlator
\begin{equation}\label{eq:boscorrbis}
\begin{aligned}
\frac{1}{|1-z|^4}\mathcal{G}^\mathrm{bos}(z,\bar z)=&\,\frac{1}{|1-z|^4}+\sum_k \frac{b_k^2}{\pi a_0^2}\Biggl[\frac{1}{|1-z|^4}\left(2 (1+|z|^2) \hat D_{3311}-\pi\right) -\frac{1}{2} \hat D_{2222}\\
&\,+\sum_{p=2}^k \frac{1}{p}\,\partial \bar \partial (|z|^2 \hat D_{p p 22})\Biggr]\,.
\end{aligned}
\end{equation}
The first line can be rewritten in a more suggestive form by making use of the identity
\begin{equation}
\begin{aligned}
\frac{1}{|1-z|^4}\left(2 (1+|z|^2) \hat D_{3311}-\pi\right)-\frac{1}{2} \hat D_{2222}= \partial \bar \partial \left[-\frac{\pi}{2}\frac{1}{|1-z|^2} + |z|^2 \hat D_{1122}\right]\,,
\end{aligned}
\end{equation}
that can be verified explicitly as in \eqref{eq: identity}. Substituting this identity in \eqref{eq:boscorrbis} one arrives at \eqref{eq:boscorr1}.

\providecommand{\href}[2]{#2}\begingroup\raggedright\endgroup


\end{document}